\documentclass[12pt]{article}

\usepackage{amsmath}
\usepackage{graphicx}
\usepackage{eufrak}
\usepackage{cite}

\renewcommand{\theequation}{\arabic{section}.\arabic{equation}}
\newcommand{\mys}[1]{\section{#1} \setcounter{equation}{0}}

\newcommand{\myappendix}{\appendix
   \renewcommand{\theequation}{\Alph{section}.\arabic{equation}}
   \vspace{30pt} \noindent {\Large \bf Appendices}}

\newlength{\dummysp}
\newcommand{\diag}{\mathop{{\hbox{diag} \, }}\nolimits}
\newcommand{\tr}{\mathop{{\hbox{Tr} \, }}\nolimits}
\newcommand{\real}{\mathop{{\hbox{Re} \, }}\nolimits}

\newcommand{\stxt}[1]{\mathop{\hbox{{\scriptsize #1}}}\nolimits}
\newcommand{\bbar}[1]{{\overline{#1}}}
\newcommand{\half}{\frac{1}{2}}

\newcommand{\beq}{\begin{eqnarray}}
\newcommand{\eeq}{\end{eqnarray}}
\newcommand{\nnn}{ \nonumber \\ }
\newcommand{\p}{{\partial}}

\newcommand{\Rbf}{{{\bf R}}}

\newcommand{\e}{{\epsilon}}
\newcommand{\s}{{\sigma}}
\newcommand{\vev}[1]{{\langle #1 \rangle}}
\newcommand{\bigvev}[1]{{\left\langle #1 \right\rangle}}
\newcommand{\ord}[1]{{{\cal O}(#1)}}
\newcommand{\gappeq}{\mathrel{\rlap {\raise.5ex\hbox{$>$}}
{\lower.5ex\hbox{$\sim$}}}}
\newcommand{\lappeq}{\mathrel{\rlap{\raise.5ex\hbox{$<$}}
{\lower.5ex\hbox{$\sim$}}}}
\newcommand{\myref}[1]{(\ref{#1})}

\newcommand{\bfe}[1]{\vspace{4pt} {\bf #1 \hspace{2pt}}}

\newcommand{\ben}{\begin{enumerate}}
\newcommand{\een}{\end{enumerate}}

\newcommand{\sqtw}{\sqrt{2}}
\newcommand{\fourth}{\frac{1}{4}}
\newcommand{\sbar}{{\bar \s}}
\newcommand{\phib}{{\bar \phi}}

\newcommand{\psib}{{\bar \psi}}

\newcommand{\bit}{\begin{itemize}}
\newcommand{\eit}{\end{itemize}}

\newcommand{\Cbf}{{\bf C}}
\newcommand{\susy}{supersymmetry}
\newcommand{\susyc}{supersymmetric}
\newcommand{\obf}{{\bf 1}}
\newcommand{\nbf}{{\bf n}}
\newcommand{\mbf}{{\bf m}}
\newcommand{\kbf}{{\bf k}}
\newcommand{\xb}{{\bbar{x}}}
\newcommand{\yb}{{\bbar{y}}}
\newcommand{\ibf}{\boldsymbol{\hat \imath}}
\newcommand{\jbf}{\boldsymbol{\hat \jmath}}
\newcommand{\lbf}{\boldsymbol{\ell}}
\newcommand{\sbf}{\boldsymbol{\s}}

\newcommand{\Ncal}{{\cal N}}

\newcommand{\SSB}{{S_{\stxt{SB}}}}
\newcommand{\hphi}{{\hat \phi}}
\newcommand{\hg}{{\hat g}}
\newcommand{\hmu}{{\hat \mu}}
\newcommand{\hginv}{{{\hat g}^{-1}}}

\newcommand{\zb}{{\bbar{z}}}
\newcommand{\xdag}{x^\dagger}
\newcommand{\ydag}{y^\dagger}
\newcommand{\zdag}{z^\dagger}
\newcommand{\hx}{{\hat x}}
\newcommand{\hxd}{{\hat x}^\dagger}
\newcommand{\hxb}{{\hat \xb}}
\newcommand{\hy}{{\hat y}}

\newcommand{\Dslash}{{\not \hspace{-4pt} D} }
\newcommand{\Ocal}{{\cal O}}
\newcommand{\tmr}{t^{\mu \nu \rho}}
\newcommand{\tmn}{t^{\mu \rho \nu}}
\newcommand{\Fb}{{\bar F}}

\newcommand{\eb}{{\bar \e \hspace{2pt}}}
\newcommand{\Qb}{{\bar Q}}

\newcommand{\xib}{{\bar \xi}}

\def\[{\left [}
\def\]{\right ]}
\def\({\left (}
\def\){\right )}

\begin{document}

\begin{titlepage}

\renewcommand{\thefootnote}{\fnsymbol{footnote}}

FTPI-MINN-06/04
\hfill Feb.~3, 2006

UMN-TH-2430/06
\hfill hep-lat/0602007

\vspace{0.45in}

\begin{center}
{\bf \Large Deconstruction and other approaches to \\ \vskip 10pt
supersymmetric lattice field theories}
\end{center}

\vspace{0.15in}

\begin{center}
{\bf \large Joel Giedt\footnote{{\tt giedt@physics.umn.edu}}}
\end{center}

\vspace{0.15in}

\begin{center}
{\it Fine Theoretical Physics Institute, University of Minnesota \\
116 Church St.~S.E., Minneapolis, MN 55455 USA }
\end{center}

\vspace{0.15in}

\begin{abstract}

\end{abstract}
This report contains both a review of recent approaches to
supersymmetric lattice field theories and some new results
on the deconstruction approach.  The essential reason for
the complex phase problem of the fermion determinant is
shown to be derivative interactions that are not present
in the continuum.  These
irrelevant operators violate the self-conjugacy of 
the fermion action that is present in the continuum.
It is explained why this complex phase problem does not
disappear in the continuum limit.
The fermion determinant suppression of various branches of
the classical moduli space is explored, and
found to be supportive of previous claims
regarding the continuum limit.
\end{titlepage}

\renewcommand{\thefootnote}{\arabic{footnote}}
\setcounter{footnote}{0}

\tableofcontents

\mys{Introduction}
Several motivations exist for efforts to 
formulate supersymmetric field theories on a lattice.
These will be summarized in \S\ref{moti}.  It
is difficult to formulate these theories in
such a way as to avoid fine-tuning of counterterms.
This is briefly discussed in \S\ref{diff}.  The points
that I discuss in \S\ref{moti}-\S\ref{diff} are all well-known; experts
may prefer to skip directly to \S\ref{summ}.  There I briefly
summarize the content of this report, with an
emphasis on new results that have not appeared
elsewhere.

\subsection{Motivations}
\label{moti}
One is just to furnish a consistent
nonperturbative definition for a given
supersymmetric field theory.
A quantum theory
that is well-defined in perturbation theory
may nevertheless have a nonperturbative
anomaly.  A classic example is, for example, the {\it Witten
anomaly} \cite{Witten:1982fp}.  A more recent, exotic example
is \cite{Hayakawa:2006fd}.
With respect to
functional integrals that appear in quantum field theory,
the lattice approach provides a rigorous, nonperturbative 
way to give them concrete meaning, and to study their behavior.
If there is a nonperturbative anomaly associated
with one of the perturbative symmetries of the continuum
theory, it would be interesting to see how that surfaces
in the lattice theory.  It has even been argued that
it is possible to have a supersymmetry anomaly \cite{Casher:1999mz},
which would be particularly interesting to study in
the lattice formulation of such theories.

A second motivation is that some 
nonperturbative aspects of supersymmetric field
theories are not calculable (in a quantitative sense) by the usual techniques.
One prime example occurs in cases involving
{\it chiral superfields.}  So-called {\it F-term}
parts of the action ({\it holomorphic data}) 
are necessarily obtained from a {\it superpotential,} 
which is a holomorphic
function of the chiral superfields.\footnote{In this discussion,
the action may include source terms for elementary
fields or composite operators.  In this case,
holomorphic data may include composites than
can be viewed as source terms in the superpotential.
Equivalently, a {\it superspace} description of
Green functions could be used; 
Green functions that are F-terms then comprise
holomorphic data.  See for example Chs.~9-10
of \cite{Wess:1992cp}.}  This part of
the action is typically subject to powerful
nonrenormalization theorems.  Together with
symmetry constraints, this often allows one
to draw conclusions about the effective superpotential;
the validity of these conclusions sometimes
extends beyond perturbation theory.  However,
parts of the action that have a nonholomorphic
origin, so-called {\it D-term} contributions 
({\it nonholomorphic data}),
can undergo significant renormalization when
running couplings turn strong.  
Quite often, the continuum methods that are
available (e.g., instanton calculus and $\Rbf^3 \times S^1$ formulations) 
are not able to make definitive
statements about this renormalization.
As a consequence,
quantitative or even qualitative features
of the low energy effective theory are not
under theoretical control, to the extent
that they depend on nonholomorphic data.
In phenomenological applications, this lack
of control over nonholomorphic data is
distressing, since the so-called {\it supersymmetry-breaking
soft-terms} that partly determine superpartner spectra
and couplings will depend on parts of the
effective action that involve nonholomorphic
quantities, such as the {\it K\"ahler potential.}
See for example \cite{Kaplunovsky:1993rd,Brignole:1993dj}.
Lattice Monte Carlo simulations would, as a first step,
give us a handle on the spectrum of light states.
Through the introduction of source fields, it
might even be possible to constrain the leading terms
in the effective K\"ahler potential, when the lattice
data is combined with what is known about
the effective superpotential by continuum methods.
This is because both the K\"ahler potential $K$ and superpotential $W$
play a role in the scalar potential: $V=K^{k \bar \ell} W_k \bar W_{\bar \ell}$,
where $K^{k \bar \ell}$ is the inverse of the K\"ahler metric $K_{k \bar \ell}
= \p^2 K/ \p \phi^k \p \bar\phi^{\bar \ell}$ and $W_k = \p W/\p \phi_k$,
$\bar W_{\bar \ell} = (W_\ell)^*$.  Thus, lattice
simulations have the potential to teach us something
about nonperturbative renormalization of nonholomorphic quantities.

A third motivation, which is closely related
to the second, is that we would like to
improve our understanding of
dynamical supersymmetry breaking.
4d supersymmetric gauge theories play
an important role in phenomenological
applications.  A strongly coupled sector
is often invoked in the mechanism that
splits the superpartners from the observed
Standard Model spectrum.  As examples, see \cite{Nilles:2004zg}
and references therein for a review of {\it supergravity mediation}
of {\it gaugino condensation}, and
see \cite{Giudice:1998bp} for a review of the {\it gauge mediation}
scenario.
Any improvement of our understanding of the strongly
coupled gauge theories that are invoked in these proposals
would be helpful.
If a good formulation of super-Yang-Mills (SYM)
on the lattice is discovered, it is likely that
interesting, well-posed questions could be studied
through lattice simulations.

A fourth motivation is that for certain supersymmetric
theories, quite a lot is known precisely, and it would
be instructive---for the development of lattice
theory and techniques---to reproduce these results on the lattice,
in simulations or by analytic methods.  For example, it would
be very nice to reproduce the exact mass formulae, confinement
and chiral symmetry breaking of pure $\Ncal=2$ SYM that
was derived by Seiberg and Witten in \cite{Seiberg:1994rs,Seiberg:1994aj}.

A fifth motivation has to do with the
evolving understanding of the relationship
between SYM and theories of quantum gravity;
in particular, string/M-theory.
A nonperturbative formulation of
superstring theory in general backgrounds is
still lacking, in spite of the recent successes
of M(atrix) theory \cite{Banks:1996vh,Ishibashi:1996xs}, 
the AdS/CFT correspondence \cite{Maldacena:1997re,Gubser:1998bc,Witten:1998qj},
the PP-wave limit \cite{Metsaev:2001bj,Metsaev:2002re,Blau:2001ne,Berenstein:2002jq}, 
etc.  However, through the studies just cited, 
nonperturbative formulations
in special backgrounds have emerged, and it is
of considerable interest to study them in
relation to SYM on the lattice for various numbers
of spacetime dimensions.  In particular,
both the Matrix theory formulations of string/M-theory,
and the AdS/CFT correspondence,
are expressed in terms of quantum theories of
dimensionally reduced SYM.  The nontrivial
vacuum dynamics takes on a gravitational
meaning.

\subsection{The difficulty}
\label{diff}
The difficulty in formulating supersymmetric
field theories is well-known, and was pointed
out long ago by Dondi and Nicolai \cite{Dondi:1976tx}.
The supersymmetry algebra fits into the larger,
super-Poincar\'e algebra.  In particular, the
supersymmetry algebra closes on the generator
of infinitesmal spacetime translation.  Since
the spacetime translation group is explicitly
broken by the discretization, the supersymmetry
algebra invariably must be modified on the lattice.
The obvious option is to have it close on discrete
translations.  However, the Leibnitz rule does
not hold on the lattice.  A result of this is
that even if the supersymmetry algebra holds
on linear expressions of lattice fields, it is
violated for more general polynomials.  It then
follows that interacting supersymmetric actions
will not be invariant w.r.t.~the lattice supersymmetry.

Since the violation of the Leibnitz rule is
an $\ord{a}$ effect, where $a$ is the lattice
spacing, the non-invariance of
the lattice action w.r.t.~lattice supersymmetry
is likewise $\ord{a}$.  Classically this
disappears in the continuum limit.  Unfortunately,
in the quantum theory the $\ord{a}$ violations
can be overcome by UV divergences.  Essentially,
any non-irrelevant operator that is allowed
by the symmetries of the lattice action will
be radiatively generated.  The lack of
exact supersymmetry in the lattice theory
generically leads to supersymmetry violating
operators in the renormalized theory.  These
can be subtracted off with counterterms.  However,
this calls for a fine-tuning of the counterterm
coefficients in order to achieve the desired
continuum limit.  A nonpeturbative method
of fine-tuning is required if one wishes to
go beyond perturbation theory.

Symmetries of the lattice theory can
sometimes overcome this difficulty.  This
approach is discussed in much more detail
below.  In particular, see \S\ref{s:vap}
below.

\subsection{Summary}
\label{summ}
One method of building lattice models with some
exact supersymmetry is based on {\it dimensional deconstruction}
\cite{Arkani-Hamed:2001ca,Hill:2000mu}.
In \S\ref{ql22}-\S\ref{ql44} I will focus on this
approach as it applies to 2d SYM with {\it extended}
supersymmetry; furthermore, I
will concentrate on the formulation that leads to
a Euclidean action formulation of the lattice theory,
as opposed to a Hamiltonian formulation, where
time is left as a continuous variable.  The 2d models are
convenient in that they provide
a simple context to illustrate the methods involved.
Furthermore, these 2d SYM models are relevant to
the ${\rm AdS}_3/{\rm CFT}_2$ correspondence---an
extension of the original ${\rm AdS}_5/{\rm CFT}_4$
ideas.  (Here, subscripts indicate spacetime
dimensions.)
Generalizations of the deconstruction
approach to higher dimensions will only be
discussed parenthetically, since the techniques
are the same as in the 2d case.

The 2d examples already illustrate the principal point:
sufficient cleverness in constucting the lattice formulation
can overcome the apparent difficulties in obtaining
the desired continuum limit.  Unfortunately, UV
behavior gets worse for $d > 2$.  This tends
to make the avoidance of unwanted renormalizations---i.e., those
that produce supersymmetry breaking non-irrelevant operators---more 
challenging, especially in the case of
most interest, $d=4$.  For $d > 4$, 
a lattice SYM formulation would only be useful as
an effective theory that captures nonperturbative
effects with a scale far below the cutoff, since
these theories are nonrenormalizable.\footnote{In
this regard, the situation would be analogous to lattice heavy
quark effective theory or lattice nonrelativistic QCD.}
Due to infrared freedom in
these theories, continuum
methods are already reliable in this regime---although
lattice analysis could nevertheless prove
instructive.  $d>4$ theories
probably require something more along the lines
of string/M-theory in order to make sense as anything
other than a low-momentum effective theory.

The deconstructed spacetime lattice constructions are all 
arrived at by {\it orbifold projections}
of SYM matrix models; i.e., in each case we quotient a
0d matrix model (SYM with all spacetime dimensions reduced out) 
by some discrete symmetry group of the
theory.  Degrees of freedom that are not
invariant with respect to the orbifold group are projected out.
This approach was introduced for a Euclidean spacetime lattice
formulation\footnote{Kaplan et al.~had previously
formulated a spatial lattice by a similar method
\cite{Kaplan:2002wv}.} by Cohen et al., 
whom I refer to hereafter as CKKU \cite{Cohen:2003xe}.
In this report I will examine both the (2,2) and (4,4)
2d SYM constructions.\footnote{
Here and below, (m,n) refers to m left-chirality and n right-chirality
supercurrents (equivalently, m left-moving and n right-moving supercurrents).
See Appendix~\ref{mnnot}.}

I now summarize the remainder of this report:
\bit
\item
Before delving into the details of the CKKU models,
I first provide in \S\ref{s:vap} a summary of various approaches to
\susy\ on the lattice, both non-gauge and gauge models.
\item
In \S\ref{ql22} I discuss aspects of the (2,2) 2d SYM
construction of CKKU \cite{Cohen:2003xe}.  In particular,
I describe how the phase of the fermion determinant
is generically complex, in a boson configuration
dependent way \cite{Giedt:2003ve}.  
In this report I present a new, more fundamental
understanding of this problem.  The origin of this phase is shown
to be $\ord{a}$ irrelevant, lattice derivative 
operators that violate
the self-conjugacy that is enjoyed by 
the continuum fermion action.
Due to the presence of modes with $\ord{a^{-1}}$
momenta in the fermion determinant, these
derivative operators make a non-negligible
contribution, even in the $a \to 0$ limit.
\item
In \S\ref{ql44} I delve into the (4,4) 2d SYM construction
of CKKU \cite{Cohen:2003qw}.  In addition 
to other matters, a discussion is given of
the classical moduli space of the lattice theory,
and the effects of the magnitude of the fermion determinant
for fluctuations about points on this moduli space.
Although some part of this discussion has
appeared elsewhere \cite{Giedt:2004tn}, a more thorough analysis
is contained here---with results that were not
previously noted.  Some simulation results in the phase-quenched ensemble
are briefly described, including my attempts to
reweight by the phase of the fermion determinant---which
turned out to be unsuccessful.  It is shown, however,
that phase-quenched averages support the
general picture of how the continuum limit
emerges.
\item
A brief set of conclusions is contained in \S\ref{s:con}.
I also summarize related research that I now have in progress.
\item
App.~\ref{mnnot} describes the simplest type of
2d field theory with (2,2) supersymmetry.  This is
provided as an illustration of the basic structure of
this sort of supersymmetry, and the logic of the notation ``(2,2).''
\item
App.~\ref{abre} briefly reviews what is known
about the boundedness of the matrix partition
functions that are used in the deconstruction
approach to lattice SYM.  It is described how
noncompact flat directions of the classical action are
rendered effectively compact in the ``quantum'' theory---i.e.,
when integration over modes orthogonal
to the moduli space is performed.
\item
A result that was lacking in earlier works is
presented in App.~\ref{dfm}.  Here I rigorously prove
that the deformed fermion matrix correctly
factors out ever-present fermion zeromodes
that occur in the deconstructed models, and
gives a projection onto the orthogonal complement.
\item
In App.~\ref{scap}, a review of generalized
conjugation of fermionic theories is provided.
It is described how unitary conjugation maps may
be exploited to prove important properties
of the fermion determinant, such as reality.
\item
App.~\ref{n4ms} gives a brief proof of the
description of the $\Ncal=4$ 4d SYM moduli space,
since the same algebraic constraints must
be satisfied in (4,4) deconstructed model
that is described in \S\ref{ql44} of the main text.
\eit


\mys{Various approaches to \susy\ on the lattice}
\label{s:vap}
Having motivated the study of lattice supersymmetry,
I next give a brief survey of some of the more
promising approaches that have appeared in the
literature of late.

We know from experience with other continuum
symmetries (e.g., Euclidean rotation invariance and
chiral symmetry) that exact lattice symmetries
are often the key to obtaining the desired
continuum limit.  This is because
the exact lattice symmetries forbid
unwanted operators:  those that are non-irrelevant
(according to naive power-counting) and would otherwise destroy
the desired symmetry in the quantum continuum limit,
absent fine-tuning of counterterms.
Thus, exact lattice symmetries can in some cases
guarantee that the long-distance Wilson effective action
will have the desired continuum symmetries.

Indeed, in some cases it has been shown how to
construct lattice actions whose symmetry group
forbids unwanted supersymmetry-breaking operators.  In most cases, this occurs due to
an exact lattice version of some part of the
continuum supersymmetry algebra, in conjunction with
other lattice symmetries.
In some cases,
there is only a partial success, in that the
number of counterterms that must be adjusted
is reduced but not zero.  In that
case, some tuning of bare parameters or additional
counterterms will still
be required to achieve the desired continuum
limit.

One should not lose sight of
the fact that a good continuum limit typically
arises from the combined constraints
of all lattice symmetries---the space group
of the lattice, discrete internal symmetries,
a subalgebra of the supersymmetry, gauge symmetry, etc.
I would also like to point out that
forbidding unwanted operators may or
may not be sufficient to guarantee the
correct continuum limit.  An example from
4d phenomenology is the following.  Generic
2-Higgs Doublet models and the Minimal Supersymmetric
Standard Model differ, among other things, in one
important respect:  in the former case the strength of the quartic
coupling in the Higgs potential is arbitrary, whereas
in the latter case it is related to the gauge
couplings in a precise manner.\footnote{Here I
refer to the quartic operators that are allowed
by supersymmetry, and that arise from D-terms.}
More generally,
supersymmetric theories are typically more constrained
than the most general theory containing the
same operators and symmetries other than supersymmetry.
These additional constraints would need to be guaranteed
by the exact lattice symmetries if one is to be
entirely successful in avoiding fine-tuning
of counterterms.

I now briefly summarize the existing threads of
construction, based mostly on exact lattice symmetries.

\subsection{Deconstruction models}  
\label{demo}
These are
lattice formulations of super-Yang-Mills that follow
an approach based on {\it dimensional deconstruction}
\cite{Arkani-Hamed:2001ca,Hill:2000mu}.
In deconstruction, dimensions are replaced by
product groups $\bigotimes_{m \in \{ {\rm sites} \}} G_m$ 
connected by representations\footnote{Here,
$\{ v^{(i)} \}$ is a set of constant vectors,
composed of elementary lattice vectors.} 
${\cal R}_{m,i} = (R_m,R_{m+v^{(i)}})$
that link the factors $G_m$, just as in lattice
gauge theory; however, in deconstruction the lattice spacing
that defines the continuum limit
is determined by a background value of the
link fields.\footnote{It is amusing to
note that product group theories with this
sort of spontaneous symmetry breakdown to
a diagonal subgroup were considered a long time ago,
but without the dimensional deconstruction
interpretation \cite{Georgi:au,Halpern:1975yj}.
I thank Marty Halpern for bringing this to
my attention.}
Examples in the literature include:  a spatial lattice~\cite{Kaplan:2002wv};
Euclidean spacetime lattices \cite{Cohen:2003xe,Cohen:2003qw,Kaplan:2005ta};
and, lattices of just one spatial dimension 
\cite{Giedt:2003xr,Poppitz:2003uz}.
Note that I have not listed the constructions that
were the original motivation for dimensional deconstruction:
theories with spacetime dimension $d>4$.  That is because
such theories are nonrenormalizable and do not
have a continuum limit, in the lattice field theory sense.
Here our interest is only
in theories where the lattice cutoff can ultimately
be removed.  The deconstruction approach 
is the subject of the bulk of this report.  Fine-tuning
is very much reduced in these constructions,
and in some cases nonexistent.

\subsection{Q-exact quantum mechanics}
\label{qeqm}
In this case,
1d supersymmetric quantum mechanics 
\cite{Catterall:2000rv,Catterall:2001wx}
is obtained from a Q-exact action (i.e., $S=QX$).
Here, $Q$ is an exact lattice supercharge
that is {\it nilpotent,} $Q^2=0$, once auxiliary
fields are introduced.  
In $Q$, derivatives are realized through discrete difference
operators; with respect to a
discrete approximation of the continuum
theory superalgebra, $Q^2=0$ is a subalgebra.  It is trivial
that the action is ``supersymmetric'' (i.e., invariant w.r.t.~$Q$): $QS=Q^2X=0$.
Models in this class 
have been derived from:  a Nicolai map 
\cite{Catterall:2000rv,Catterall:2001wx},
a relationship to topological field theory 
\cite{Catterall:2003xx,Catterall:2003wd},
and superfield techniques \cite{Giedt:2004qs}.  
Fine-tuning of counterterms is absent in the continuum
limit.  This has been shown in perturbation theory
and at a nonperturbative level \cite{Giedt:2004qs}.
By contrast, in a naive discretization
a counterterm must be introduced \cite{Giedt:2004vb}.

\subsection{Q-exact 2d Wess-Zumino models}
\label{qewz}
In this case the lattice theory is supposed
to provide a Landau-Ginsburg description of
the minimal discrete series of
$\Ncal=2$ superconformal field theories.
The continuum Landau-Ginsburg effective theory
is nothing but the 4d $\to$ 2d dimensional
reduction of 4d Wess-Zumino models \cite{Wess:1973kz}.  

These Q-exact lattice actions were first formulated 
in \cite{Elitzur:1983nj,Sakai:1983dg} using 
{\it Nicolai map} \cite{Nicolai:1979nr} methods, 
relying on earlier Hamiltonian \cite{Elitzur:1982vh}
and continuum \cite{Cecotti:1982ad} studies
that also utilized the Nicolai map.  This
Nicolai map approach was also described in
\cite{Catterall:2001wx,Catterall:2001fr}.
The same class of lattice actions was obtained
by superfield techniques in \cite{Giedt:2004qs}.
Once auxiliary fields are introduced, the
lattice action takes a Q-exact form,
as was emphasized in topological
interpretation of \cite{Catterall:2003xx,Catterall:2003wd}
and the lattice superfield approach of \cite{Giedt:2004qs}.  
Here again, there is a nilpotent subalgebra $Q^2=0$ that
is preserved in the formulation with auxiliary
fields.

Detailed studies of the spacetime lattice system
have been performed: in \cite{Beccaria:1998vi} by stochastic
quantization methods; in \cite{Catterall:2001wx,Giedt:2005ae}
by the Monte Carlo simulation approach; and in \cite{Giedt:2004qs}
by lattice perturbation theory.

It has been shown in \cite{Giedt:2004qs} that 
the continuum limit
of the lattice perturbation series is identical
to that of the continuum theory,
due to cancellations that follow from $Q^2=0$.
Thus, the Q-exact spacetime lattice has
behavior that is similar to what was
found on the $Q,Q^\dagger$-preserving
spatial lattice in \cite{Elitzur:1983nj}.
However, it was also shown in \cite{Giedt:2004qs}
that the most general continuum
effective action that is consistent
with the symmetries of the bare lattice action
is not the (2,2) 2d Wess-Zumino model.
Thus the nonperturbative studies 
\cite{Beccaria:1998vi,Catterall:2001wx,Giedt:2005ae} cited above
are important to settling the question
as to whether or not the correct
continuum limit is obtained.

The spectrum degeneracy and approximate satisfaction
of the supersymmetry Ward-Takahashi (WT) identities that was observed
in the studies \cite{Beccaria:1998vi,Catterall:2001wx} indicate that
the correct continuum limit is obtained
without fine-tuning, even at a nonperturbative
level.  In \cite{Giedt:2005ae} the R-symmetries of this lattice system
were studied.  It was
found that the continuum symmetries do seem
to be recovered in the nonperturbative,
quantum continuum limit.  In summary, all indications
are that the lattice system has the correct
continuum limit without any fine-tuning.

\subsection{Supersymmetric nonlinear sigma models}
Supersymmetric 2d nonlinear sigma models have been attempted
on the lattice.  Here both topological
\cite{Catterall:2003xx,Catterall:2003wd}
and superfield \cite{Giedt:2004qs} 
approaches have been applied.  One
finds that it is necessary to give up either
the exact nonlinear symmetry or the exact lattice
supersymmetry, assuming the nonlinear
symmetry group is nonabelian.  It remains to be seen
which is the better option and whether or not fine-tuning
of counterterms is required.  The nonlinear
symmetries of the 2d sigma model are important
to the renormalizability of the theory, so
sacrificing them could create severe problems
with the continuum limit.  
Encouraging analytic results 
and some Monte Carlo simulations of the $CP^1$ model have been reported
in \cite{Catterall:2003uf,Ghadab:2004rt}.

\subsection{Twisted geometrical SYM}
These are formulations that rely on the
geometrical picture that relates
twisting of spinors and the K\"ahler-Dirac
formalism.  With the scalar supercharge $Q$ 
that is exposed in this procedure, and
all elementary degrees of freedom encoded
in K\"ahler-Dirac fields, Catterall has been
able to construct Q-exact actions for super-Yang-Mills (SYM)
with extended supersymmetry.
The ideas behind Catterall's construction
go back to his topological approach for non-gauge theories,
described in \cite{Catterall:2003xx,Catterall:2003wd}.  This is because
twisting is an integral part of the
formulation of topological field theories.
In \cite{Catterall:2004np} the $\Ncal=2$ 4d SYM
was formulated, whereas in \cite{Catterall:2005fd}
the $\Ncal=4$ 4d SYM was constructed.

\subsection{Q-exact compact super-Yang-Mills}
\label{qecs}
These are the models of 
Sugino~\cite{Sugino:2003yb,Sugino:2004qd,Sugino:2004uv,Sugino:2006uf}, 
where his approach is to modify the supersymmetry
transformation in such a way that it has a simple
action on the link fields that appear in a 
compact formulation of SYM.  
He achieves this by supplementing the
lattice supersymmetry tranformation with
$\ord{a}$-suppressed operators.
He then defines the action as a
Q-exact quantity, where $Q^2=0$ up
to a gauge transformation.  By this approach he is
able to construct a variety of extended
SYM theories with an exact lattice
supersymmetry.  He demonstrates the absence
of fine-tuning at the perturbative level,
in way that is quite similar to what was done in the deconstruction
models.  Recently, there appeared a brief note
that questioned Sugino's claim that he has been able
to obtain a unique classical vacuum about
which to define the continuum limit \cite{Chu:2005ps}.
However, this note is wrong in its conclusions,
due to several errors that are so elementary that
I need not detail them here.

My own studies have found that the fermion action
is not hermitian, due to irrelevant operators,
indicating a complex fermion determinant.
I will present these results in a forthcoming
article.

\subsection{Perfect action models}
Direct constructions in the spirit
of the Ginsparg-Wilson relation, as 
suggested by L\"uscher \cite{Luscher:1998pq},
have been considered in a few articles 
\cite{Aoyama:1998in,Bietenholz:1998qq,So:1998ya}.
However, the only cases where this
has been successfully applied are theories with quadratic
actions; i.e., free theories.  
Quadratic theories can be successfully latticized
in a naive approach, if a judicious choice of discrete
difference operators is made.  Furthermore, a
lattice approach is not needed for quadratic
theories, since they can be solved exactly
in the continuum.  Thus, the 
perfect action approach has not yet
proven useful for interesting (i.e., interacting) 
supersymmetric field theories on the lattice.

\subsection{Super-renormalizable lattice field theories}
In \cite{Golterman:1988ta}, Golterman and Petcher studied
the $\Ncal=1$ 2d Wess-Zumino model.  
They showed by
Reisz lattice power-counting \cite{Reisz:1987da} that
the lattice perturbation theory is super-renormalizable,
with all diagrams of UV degree $D \geq 0$ appearing
at 1-loop.  Thus, the counterterms that
were needed to guarantee the perturbative continuum
limit could all be exactly
determined by a straightforward 1-loop computation.

In the proof that the supersymmetry WT identities
were respected in the continuum limit, Golterman
and Petcher exploited a modified lattice supersymmetry
algebra that contained a series of higher dimensional
operators, suppressed by positive powers of the
lattice spacing.  This sort of
nonlocal, exact lattice supersymmetry was
inspired by the earlier work of Banks and Windey \cite{Banks:1982ut}.
The Golterman and Petcher method of analysis inspired
the {\it nonlinear Q models} that will
be discussed in \S\ref{nlqm} below.

More generally,
consider the case of any lattice field theory that is
super-renormaliz\-able according to the Reisz
power-counting rules.  It follows that such
a theory allows for a counterterm adjustment 
that will guarantee the correct perturbative continuum
limit.  Exactly this sort of approach has been 
carried out for 3d $\Ncal=2$ SYM \cite{Elliott:2005bd}.
Another interesting result is \cite{Suzuki:2005dx},
where (2,2) 2d SYM is formulated with chiral
lattice fermions (cf.~\S\ref{chif}), so that the only counterterms
are scalar masses.  These are entirely determined at 1-loop
since the theory is super-renormalizable.

However, the crucial role of the
lattice is supposed to be as a nonperturbative
definition of the quantum field theory.
One must ask:  To what extent does good perturbative
behavior lead to good nonperturbative
behavior?  Here, there is only anecdotal
data from a few nonperturbative analyses of
supersymmetric lattice theories; these have
been referred to above or below.
It does seem that good pertubative behavior
correlates with good nonperturbative behavior for
these few simple models; for example, adjustment
of counterterms as determined from a super-renormalizable
lattice perturbation theory has been found
to give good nonperturbative results in 
the naive latticization of supersymmetric
quantum mechanics \cite{Giedt:2004vb}.
Also, some models where it was
shown that no counterterms are required
in perturbation theory have been studied
by nonperturbative means, with the result
that the correct continuum limit appears to
emerge at the nonperturbative level as well:
the examples here are the Monte Carlo studies
of Q-exact lattice
supersymmetric quantum mechanics \cite{Giedt:2004vb} (c.f.~\S\ref{qeqm}),
and nonperturbative studies of the
Q-exact $\Ncal=2$ 2d Wess-Zumino model by
stochastic \cite{Beccaria:1998vi} 
and Monte Carlo approaches 
\cite{Catterall:2001wx,Catterall:2001fr,Giedt:2005ae} (c.f.~\S\ref{qewz}).
But without a rigorous proof, one cannot
say in what cases these types of positive results
will persist.  The best one can do is
to test each theory nonperturbatively; say,
by checking supersymmetric WT identities.

\subsection{Nonlinear Q models}
\label{nlqm}
The work of Goltermann and
Petcher \cite{Golterman:1988ta} has
recently inspired Bonini and Feo to formulate a 
nonlocal, exact lattice supersymmetry 
for a discretized version of the 4d Wess-Zumino action 
\cite{Bonini:2004pm,Bonini:2005qx,Bonini:2004wv}.  This action makes use
of Ginsparg-Wilson fermions, so that the $U(1)_R$
symmetry of the massless theory is respected without
fine-tuning \cite{Fujikawa:2001ka,Fujikawa:2001ns,Fujikawa:2002ic}.
While the nonlocal exact lattice supersymmetry
proposed by Bonini and Feo is a symmetry of the partition
function, order by order, it remains to
be seen what it has to do with symmetries
of the tranfer matrix, and hence Hamiltonian.  
Note that an operator that is nonlocal in
the time direction touches all
time-slices.  As a consequence, if this
operator generates a symmetry of
the partition function, it does not necessarily imply that
the operator commutes with the transfer
matrix.  If the operator could be shown
to localize in the continuum limit, then
it would indeed commute with the coarse-grained Hamiltonian
in this limit.  I am currently investigating these matters.

The analysis of the continuum
limit of perturbative lattice WT identities
that was reported by Bonini and Feo in \cite{Bonini:2005qx} indicates
that the exact nonlocal lattice symmetry is
related to a symmetry of the continuum Hamiltonian
through $\ord{g^2}$.  However, according to Kikukawa and Susuki \cite{Kikukawa:2004dd},
Ref.~\cite{Fujikawa:2001ka} shows that there is
already supersymmetry breaking at 1-loop order in
the wavefunction renormalization.  It is unclear to
me how this can be reconciled with the findings of
Bonini and Feo regarding the WT identities.

\subsection{Chiral fermion models of 4d $\Ncal=1$ SYM}
\label{chif}
Here a fair amount of work has been done.
It began with the observation of Curci and Veneziano
that for 4d pure $\Ncal=1$ SYM, the violation
of the \susy\ WT identity is proportional to
the mass-violation of the axial $U(1)_A$ anomalous 
WT identity \cite{Curci:1986sm}.
This is easy to understand:  if one assumes gauge 
and Lorentz invariance, the only non-irrelevant operator
that can break supersymmetry for 4d $\Ncal=1$ SYM is
a gluino mass.\footnote{The gluino is the fermionic
superpartner to the gluon.}

Thus the supersymmetric limit coincides with
the chiral limit of the lattice theory; that is,
the limit of vanishing renormalized mass.
Due to the axial anomaly, in the case of $SU(N)$ SYM,
only a $Z_{2N}$ subgroup
of $U(1)_A$ is a symmetry of the partition function.
This is spontaneously broken to $Z_{2}$ due
to condensation of the gluino bilinear, $\vev{\lambda \lambda}
\not= 0$ in a 2-component Weyl spinor notation.
Consequently, one can either (1) fine-tune the bare
mass for a lattice fermion that explicitly breaks
the $Z_{2N}$ chiral symmetry, or (2) use a chiral fermion
formulation---where there is only multiplicative
renormalization, and setting the bare mass to
zero will correspond to the chiral limit.  

In case (1), Wilson fermions have
been explored extensively (see for example
\cite{Farchioni:2001wx,Montvay:2001aj,Feo:2002yi} and references therein).  
The chiral limit corresponds to a first order phase
transition that occurs as one varies the bare mass
such that one passes through the critical mass
where $N$ vacua coexist.  See the review \cite{Montvay:2001aj}
for an extensive discussion of this matter
and recent findings in numerical simulations.

In case (2), several
formulations have been considered.  Staggered 
(or, Kogut-Susskind) fermions 
\cite{Kogut:1974ag,Susskind:1976jm,Sharatchandra:1981si},
which possess symmetries that forbid additive
mass renormalization, were studied in \cite{Kogut:1986jt}.
Because of a 4-fold increase in modes, and
because SYM has Majorana fermions, leading to
a Pfaffian,\footnote{Up to a 2-fold branch ambiguity, the
Pfaffian of a matrix is the square root of its determinant.
The branch may be determined from the precise definition
of the Pfaffian, which is well-known and I will not
repeat here.  In the cases where the Pfaffian does not have constant
complex phase (or sign, if it is real), the determination of and
sample weighting by
this factor often inhibits numerical studies.}
an eighth-root trick was applied to the staggered
fermion determinant.  This technique is 
controversial, so other formulations are of
interest, as a cross-check.  

Of particular interest are chiral formulations
that are collectively referred to as Ginsparg-Wilson
fermions, since they satisfy the Ginsparg-Wilson
relation \cite{Ginsparg:1981bj}, 
a lattice variant of chiral symmetry \cite{Luscher:1998pq}.
Domain wall fermions \cite{Kaplan:1992bt}
yield a chiral formulation in the limit of an
infinite 5th dimension.
Monte Carlo simulations of $\Ncal=1$ 4d SYM have been performed
for domain wall fermions by Fleming 
et al.~\cite{Fleming:2000fa}.  The cost of
such simulations is high, so only limited results
have been obtained.  The gluino condensate was
studied; however, the data that was obtained was not sufficient
to extrapolate to the continuum limit.  
The {\it overlap determinant} gives an
exact representation of the Weyl determinant
in the continuum limit 
\cite{Narayanan:1992wx,Narayanan:1993sk}.  
Since in 4d a Weyl fermion
is equivalent to a massless Majorana fermion,
the overlap determinant is guaranteed to yield
the correct continuum theory.
A lattice Dirac fermion that enjoys exact
chiral symmetry was later obtained by Neuberger,
based on the overlap Weyl determinant 
\cite{Neuberger:1997fp,Neuberger:1998wv},
and this is what is commonly referred to as an
overlap fermion; it is constructed from the
Wilson-Dirac operator.  In the case of the overlap
formulation, only analytical studies have 
been performed \cite{Maru:1997kh},
due to the fact that simulations are extraordinarily
expensive; significant supercomputer resources
would be required to obtain meaningful results
for the 4d $\Ncal=1$ SYM theory.

\subsection{Chiral fermion formulations of the 4d Wess-Zumino \\
model}
Some of the efforts in this direction, using
Ginsparg-Wilson fermions, were mentioned in \S\ref{nlqm} above.
Such models were first described by Fujikawa and Ishibashi in
\cite{Fujikawa:2001ka,Fujikawa:2001ns,Fujikawa:2002ic}.  The analysis of
that construction by Bonini and Feo was already summarized in \S\ref{nlqm}.
A modified formulation was given by Kikukawa and 
Susuki \cite{Kikukawa:2004dd}.  In their constuction
they have shown that the lattice chiral symmetry
reduces the number of counterterms that are required,
but does not eliminate them.  In order to obtain the
correct continuum limit, a tuning must be performed
at each order in perturbation theory.  In light of
this, the seemingly contradictory claims of Bonini and Feo regarding the
Fujikawa and Ishibashi construction need to be re-examined.

\subsection{Twisted superspace models}
In \cite{D'Adda:2004jb,D'Adda:2005zk}, models have been
introduced in which superspace is given a noncommutative,
lattice-geometrical interpretation.  For example, the
supercharges correspond to a sort of ``square-root'' of
lattice translation.  Key in this construction is
the K\"ahler-Dirac formalism and complexification
of the gauge group.  Studies of dynamical
properties of these lattice systems have not yet
been performed.

\subsection{Brute force models}
Here one just identifies the most general renormalizable
lattice action that contains the \susyc\ target, and
fine-tunes the bare parameters to obtain the desired
theory.  An example is the work of Montvay for 4d
$\Ncal=2$ SU(2) SYM \cite{Montvay:1994ze}.  There the
most general theory consistent with gauge invariance
and parity (easily preserved on
the lattice) is an $SU(2)$ adjoint Higgs-Yukawa theory,
which he formulates with Wilson fermions and compact
gauge fields (link variables).  He then explores perturbative RG flows
and identifies the trajectory that gives the desired
theory.  A nonperturbative extension of this work
would be very interesting, due to the existence
of exact results in the continuum \cite{Seiberg:1994aj}.

\subsection{Transverse lattice models}
In this class of models, lattice techniques are
combined with light-cone quantization.  The dimensions
transverse to the light-cone are formulated on a lattice.
This allows for more symmetry to be preserved (associated
with the light-cone), and reduces the dimensionality
of the lattice system that must be simulated.  An
example of this type of construction is \cite{Harada:2004wu}.

\subsection{Hamiltonian approaches}
In this case, time is treated continuously,
whereas spatial dimensions are discretized.
This allows the part of the \susy\ algebra
that closes on the hamiltonian to be realized
exactly.
A wealth of formulations and studies along
these lines can be found in the literature.
As examples, I cite \cite{Elitzur:1982vh,Elitzur:1983nj,
Kirchberg:2004vm,Beccaria:2004ds,Beccaria:2001tk,
Beccaria:2001qm,Li:1995ip,Schiller:1984cx,Ranft:1984zs,
Ranft:1983ag}.  See also
references in \cite{Feo:2002yi}.
For $d>2$ field
theories, hamiltonian formulations have traditionally
met with only limited success.

\mys{Deconstructed (2,2) 2d SYM}
\label{ql22}
In this section one of the deconstruction models
of CKKU will be discussed \cite{Cohen:2003xe}.
To lay a groundwork for the discussion of the lattice
theory, I begin with a brief review of the continuum,
or, {\it target} theory.  For the most part this
discussion will be classical, merely defining the
action and the fields that appear in it.

\subsection{The continuum theory}
The theory that I discuss is (2,2) 2d SYM.
The ``(2,2)'' denotes the supersymmetry of the theory,
generated by supercharges $Q_\pm, \Qb_\pm$.
The supercharges comprise two left-handed
2d Weyl spinors, $Q_-,\Qb_-$, 
and two right-handed 2d Weyl spinors, $Q_+,\Qb_+$.
If equations of motion are imposed, one finds that 
the notation ``(2,2)'' equivalently
denotes the number of left-moving (depending on $t+x$) 
and right-moving (depending on $t-x$) 2d supercurrents;
see Appendix~\ref{mnnot}.
The supercharges are derived from the supercurrents
in the usual way.  Often (2,2) supersymmetry is
denoted as $\Ncal=2$ \susy.  However, I prefer
the former notation since it is possible to build
chiral models; e.g., models with (0,2) \susy.  As an
example, generic 10d $\to$ 4d heterotic sting compactifications
with $\Ncal=1$ 4d target space \susy\ have (0,2) worldsheet
\susy.  Compactifications with (2,2) worldsheet supersymmetry
are the exception, corresponding to so-called {\it standard
embeddings.}

The (2,2) 2d SYM theory is most easily obtained as 
the dimensional reduction of $\Ncal=1$ 4d 
SYM to 2d.  Note that $\Ncal=1$ 4d SYM is nothing but
4d YM with a single massless adjoint Majorana
fermion.  It is minimal,
in the sense that the (four) supercharges organize themselves
into just one 4d Majorana spinor.  4d
theories of extended supersymmetry,
where $\Ncal > 1$, have more than four supercharges,
corresponding to more than one Majorana spinor.  
The dimensional reduction to 2d is the naive
one:  compactify two
dimensions on a torus, and truncate to just the zeromodes
in those dimensions.
Equivalently, impose that fields only depend on
two coordinates $t,x$.  Then 
\beq
\frac{1}{g_4^2} \int dt dx d^2y (\cdots) \to
\frac{V_2}{g_4^2} \int dt dx  (\cdots),
\eeq
where $V_2$ is the volume of the 2-torus.
Consequently the 2d coupling is $g_2^2 = g_4^2/V_2$.
It follows that $[g_2]=1$, where $[\cdots]$ indicates
mass dimensions throughout this report.
From the 2d point of view,
the two vector-boson modes along the
compactified dimensions are considered as real scalars
in the adjoint representation.  Thus, the
4d field-strength decomposes as
\beq
F_{\mu \nu} \to \{ F_{ij}, \; D_i s_a, \; [s_a,s_b] \} ,
\quad \mu,\nu=0,1,2,3, \quad i=0,1, \quad a,b=1,2,
\eeq
where $s_a$ are the scalars and
$D_i s_a = \p_i s_a + i [v_i, s_a]$.
Additionally, it is convenient to define the
complex scalar $s=(s_1 + is_2)/\sqtw$.

The (real-time) action is:
\beq
S &=& \int d^2 x \; \frac{1}{g_2^2} \tr \[ - (Ds) \cdot (Ds)^\dagger
+ \sqtw i \psib_- D_+ \psi_- + \sqtw i \psib_+ D_- \psi_+
- \fourth F \cdot F \right. \nnn
&& \left. + i \sqtw \psib_- [s^\dagger,\psi_+] 
+ i \sqtw \psib_+ [s, \psi_-]
- \half [s^\dagger, s]^2 \] .
\label{22cont}
\eeq
Here, $\psi_\pm$ are 2d Weyl spinors, which
are 1-component objects.  Also,
$D_\pm = \p_\pm + i [v_\pm, \cdot ]$,
$\p_\pm = (\p_t \pm \p_x)/\sqtw$ and $v_\pm = (v_t \pm v_x)/\sqtw$.
The field strength, $F_{01} = \p_t v_x - \p_x v_t + i [v_t, v_x]
\equiv E$, is just the color-electric field, since we are in 2d.
This action contains a chiral $U(1)_R$ symmetry:\footnote{The
subscript ``R'' is traditional, denoting a flavor symmetry
that does not commute with \susy.}
$\psi_L \to e^{i\varphi} \psi_L$, $\psi_R \to e^{-i\varphi} \psi_R$,
$s \to e^{2i\varphi} s$.  
The (2,2) supersymmetry of the action is
\beq
&& \delta v_- = i \sqtw ( \psib_+ \xi_+ - \xib_+ \psi_+), \quad
\delta v_+ = i \sqtw ( \psib_- \xi_- - \xib_- \psi_-), \nnn
&& \delta s = i \sqtw ( \psib_- \xi_+ - \xib_- \psi_+), \qquad
\delta s^\dagger = i \sqtw ( \psib_+ \xi_- - \xib_+ \psi_-), \nnn
&& \delta \psi_- = 2 D_- s^\dagger \xi_+ + (E - i[s^\dagger,s]) \xi_-, \nnn
&& \delta \psi_+ = 2 D_+ s \xi_- - (E - i[s^\dagger,s]) \xi_+, \nnn
&& \delta \psib_- = 2 \xib_+ D_- s + \xib_- (E + i [s^\dagger, s]), \nnn
&& \delta \psib_+ = 2 \xib_- D_+ s^\dagger - \xib_+ (E + i [s^\dagger, s]),
\eeq
where $\xi_\pm,\xib_\pm$ are infinitesmal Grassmann parameters.
This is an invariance provided the equations of motion are
imposed (on-shell \susy).  If auxiliary fields are introduced,
one can avoid the use of the equations of motion (off-shell \susy).
To obtain the on-shell formulation, one must impose
the so-called {\it Wess-Zumino} gauge.  This field
redefinition eliminates auxiliary fields but leaves
the usual gauge symmetry unbroken.
The on-shell \susy\ then involves a mixture the off-shell \susy\ and
a gauge transformation.  
(For further details, the reader is referred to
\cite{Wess:1992cp}.)  

After continuation to Euclidean space,
a slight change in notation, and
the introduction of 2d Dirac $\gamma$-matrices,
the action \myref{22cont} can be written in the form:
\beq
S &=& \int d^2 x \; \frac{1}{g_2^2} \tr \[ (Ds) \cdot (Ds)^\dagger
+ i \psib \Dslash \psi + \fourth F \cdot F \right. \nnn
&& \left. + i \sqtw \( \psib_L [s,\psi_R] + \psib_R [s^\dagger, \psi_L] \)
+ \half [s^\dagger, s]^2 \].
\label{22ck}
\eeq
This imaginary-time formulation is the target of the lattice
theory.

\subsection{Construction of the daughter theory}
As summarized briefly in \S\ref{demo},
the method of building the class of models 
described here is based on deconstruction
of extra dimensions \cite{Arkani-Hamed:2001ca,Hill:2000mu}.
These lattice constructions are all arrived at by {\it orbifold projections}
of supersymmetric matrix models; i.e., in each case we quotient the
matrix model by some discrete symmetry group of the
theory.  Degrees of freedom that are not
invariant with respect to the orbifold generators are projected out.
In the case of the Euclidean spacetime lattice
formulations \cite{Cohen:2003xe,Cohen:2003qw,Kaplan:2005ta}, 
the matrix models are 0d:
they are obtained by reducing all
dimensions of a Euclidean field theory.
In the case of the Hamiltonian (spatial) lattice 
formulations \cite{Kaplan:2002wv},
the matrix models are 1d: the matrix partition
function is a functional integral corresponding
to a quantum mechanical system with continuous
time.

Here I will focus on the orbifolded supersymmetric
matrix models that define a Euclidean lattice
formulation of (2,2) 2d super-Yang-Mills with $U(k)$ gauge group.  The
{\it mother theory} is a nonorbifolded $U(kN^2)$ supersymmetric
0d matrix model.  The {\it daughter theory} is obtained by
orbifolding the mother theory by a $Z_N \times Z_N$
symmetry group.  This leaves intact (among other things) a $U(k)^{N^2}$
symmetry group.  This can be associated with
a symmetry that acts as an independent
$U(k)$ at each site of an $N \times N$ lattice.
Thus, we obtain an $N \times N$ lattice theory 
with $U(k)$ gauge symmetry.  
Classically, the continuum theory is obtained
by studying the daughter theory about a particular
minimum of the bosonic lattice action.
This aspect, and its extension to the quantum
theory, will be discussed in more detail in \S\ref{cat2} below.

The action of the mother theory is just
\beq
S = \frac{1}{g^2} \tr \( \fourth v_{mn} v_{mn} 
+ \psib \sbar_m [v_m, \psi] \), \quad m=0,\ldots,3,
\label{22mta}
\eeq
where $v_{mn}=i[v_m,v_n]$, and $\sbar_m=(1,i\s_i)$ are the
Euclidean Weyl matrices, with $\s_i \;
(i=1,2,3)$ Pauli matrices; 
later we will also make use of 
the conjugate Weyl matrices $\s_m=(1, -i \s_i)$.
Each of the quantities appearing in \myref{22mta}
is decomposed on a basis of hermitain matrices
of order $kN^2$:
\beq
v_m = v_m^\alpha T^\alpha, \quad
\psi = \psi^\alpha T^\alpha, \quad
\psib = \psib^\alpha T^\alpha, \quad 
\alpha=0,\ldots,(kN^2)^2-1 .
\eeq
This structure is inherited from the $U(kN^2)$
YM gauge theory, with adjoint Majorana fermions,
from which the reduction \myref{22mta} springs.
It is assumed that we work in a basis
where $\tr T^\alpha T^\beta = \delta^{\alpha \beta}$.

Symmetries play an important role in the
construction of the lattice theory.  Since the mother
theory is obtained from a 4d $\to$ 0d reduction,
all symmetries become global.  Thus, the remnant of
gauge symmetry is just
\beq
v_m \to U^\dagger v_m U, \quad \psi \to U^\dagger \psi U,
\quad \psib \to U^\dagger \psib U, \quad U \in U(kN^2),
\label{mtgs}
\eeq
where the matrices $U$ are taken in the $kN^2 \times kN^2$
defining representation. 

Since the gauge symmetry is now global, no problems with infinities
arise from gauge orbits (at finite $N$).  Thus, in the mother theory partition function,
there is no need for gauge fixing.  The exact
value of the $SU(kN^2)$ part of the mother theory
partition function is known \cite{Krauth:1998xh,Moore:1998et}.  
It is finite---with
a value given by the $D=4, N \to kN^2$ case of 
Eqs.~(26)-(27) of \cite{Krauth:1998xh}---in 
spite of a noncompact classical moduli space.  (Entropic
effects render the quantum moduli space compact; cf.~App.~\ref{abre}.)
The $U(1)_{{\rm diag}}$ part of the mother theory partition
function is ill-defined, as it is a massless free
theory.  It is completely decoupled from the $SU(kN^2)$
part, and is irrelevant to all the considerations that
follow, though one must be careful to factor it out,
as will be discussed.

The model also possesses an $SU(2)^2 \times U(1)$ symmetry
that is inherited from global symmetries of the 4d theory.
The symmetry transformation is:
\beq
\sbar_m v_m \to L \sbar_m v_m R^\dagger, \quad
\psib \to \psib L^\dagger e^{-i \theta}, 
\quad \psi \to e^{i \theta} R \psi,
\quad L,R \in SU(2) .
\label{22rs}
\eeq
This corresponds to the $U(1)_R$ and
the Euclidean rotation group $SO(4) \simeq SU(2)^2$
of the 4d SYM.  Finally one has the 0d
reduction of \susy.  This
I write in terms of infinitesmal 2-component
Grassmann parameters $\kappa,\bar \kappa$:
\beq
&& \delta v_m = -i \psib \sbar_m \kappa + i \bar \kappa \sbar_m \psi ,
\quad \delta \psi = \fourth v_{mn} (\s_m \sbar_n - \s_n \sbar_m)
\kappa, \nnn
&& \delta \psib = -\fourth \bar \kappa
(\sbar_m \s_n - \sbar_n \s_m) v_{mn}.
\label{stra}
\eeq
Since the mother theory is formulated without auxiliary fields,
it is necessary to use the equations of motion to obtain
the \susyc\ invariance.  In this 0d case, these are just (graded)
algebraic constraints:
\beq
0 = [v_m, v_{mn}] - i \sbar_n^{\dot \alpha \alpha}
\{ \psib_{\dot \alpha}, \psi_\alpha \}
= \sbar_m [v_m,\psi] = [v_m, \psib] \sbar_m .
\eeq

In the partition function of the matrix model,
$\psi$ and $\psib$ are
independent integration variables, 
not necessarily related by a conjugation constraint.
Consequently, the supersymmetry transformation
parameters $\kappa$ and $\bar \kappa$ that
appear in \myref{stra} can also
be taken independently.  E.g., there is (formally, at least) 
nothing inconsistent with $\kappa \not= 0, \; \bar \kappa=0$.
It is still a symmetry of $Z$.  These statements
are important below when I discuss supersymmetry
after orbifold projection.

Also important below will be the decomposition
$U(kN^2) \subset U(k)^{N^2}$.  I write
the (hermitian) generators of $U(kN^2)$ in this
basis: 
\beq
&& (T^\alpha)_{\mu,\nu,i;\mu',\nu',i'},
\qquad \mu,\nu,\mu',\nu' = 1,\ldots,N, \qquad i,i' = 1,\ldots,k,
\nnn && \alpha = 0,\ldots, (kN^2)^2-1, \qquad
T^0 = \frac{1}{\sqrt{kN^2}}
\obf_{kN^2}, \qquad \tr T^{\alpha \not= 0} = 0.
\label{tmtd}
\eeq

CKKU exploit a $Z_N^2$ subgroup of the $SU(2)^2 \times U(1)$
in \myref{22rs} for the orbifold action, and embed it
into a subgroup of $U(kN^2)$ symmetry \myref{mtgs} to
break $U(kN^2) \to U(k)^{N^2}$.
The $Z_N^2$ action is generated by
$\exp(2\pi i r_a /N)$ with $N$-alities $r_a \; (a=1,2)$
that I now describe.  

First we define the combinations
\beq
x=\frac{v_0 - i v_3}{\sqtw}, \quad
y = \frac{-v_2 -i v_1}{\sqtw}
\label{xyd}
\eeq
and fermion components
\beq
\psi = \binom{\lambda}{\xi}, \quad
\psib = (\alpha , \beta ) .
\eeq
From \myref{xyd}, one finds that
\beq
v = v_m \sbar_m = v_0 + i {\bf v \cdot} \sbf = \sqtw
\begin{pmatrix} \xdag & -y \cr \ydag & x \cr \end{pmatrix} .
\label{v2mat}
\eeq
With these expressions it is straightforward
to work out the action \myref{22mta} in terms
of the new variables.

\begin{table}
\begin{center}
\begin{tabular}{c|cccccc}
& $x$ & $y$ & $\lambda$ & $\xi$ & $\alpha$ & $\beta$
\\ \hline
$r_1$ & 1 & 0 & 0 & -1 & 1 & 0 \\
$r_2$ & 0 & 1 & 0 & -1 & 0 & 1
\\ \hline
\end{tabular}
\end{center}
\caption{$N$-alities of the fields w.r.t.~the $Z_N^2$
subgroup that CKKU select from $SU(2)^2 \times U(1)$. 
\label{nalt}}
\end{table}

The $N$-alities of these fields are given in Table \ref{nalt}.
They are generated by a diagonal subalgebra of the
$su(2)_L \oplus su(2)_R \oplus u(1)_Y$ associated with \myref{22rs}:
\beq
r_1 = -L_3 + R_3 - Y, \qquad
r_2 = L_3 + R_3 - Y,
\eeq
as can be checked.  Note that the fermion $\lambda$
is neutral.  It turns out that in the
CKKU approach the number
of fermion fields that are neutral w.r.t.~the orbifold
group is identical to the number of supersymmetries that
are left intact.  This simple rubric illustrates
the usefulness of the orbifold technique.

The $Z_N^2$ action is embedded in the gauge
group as follows.  We write
\beq
\Omega=\diag(\omega,\omega^2,\ldots,\omega^N), \qquad
\omega=\exp(2\pi i/N).
\label{omdf}
\eeq  
Then define
\beq
C_1 = \Omega \otimes \obf_N \otimes \obf_k, \qquad
C_2 = \obf_N \otimes \Omega \otimes \obf_k .
\label{c12df}
\eeq
Thus $C_1$ distinguishes only the
indices $\mu, \mu'$ in \myref{tmtd}
and $C_2$ distinguishes only the
indices $\nu, \nu'$ in \myref{tmtd}.
The action of the orbifold group (with embedding)
on any of the fields in Table \ref{nalt} is ($a=1,2$):
\beq
\Phi \to e^{2 \pi i r_a / N} C_a \Phi C_a^\dagger .
\label{rcpc}
\eeq
Keeping only the components that are
neutral under this orbifold action corresponds to
the ``orbifolded'' theory, which will be referred to
as the {\it daughter theory,} in keeping with
the appelations of CKKU \cite{Cohen:2003xe}.  I now examine
the explicit form of the daughter theory action.

It is useful to be very explicit about
\myref{rcpc}.  Using the notation of \myref{tmtd},
each field in Table \ref{nalt} can be written in
the form $\Phi_{\mu,\nu,i;\mu',\nu',i'}
= \Phi^\alpha (T^\alpha)_{\mu,\nu,i;\mu',\nu',i'}$.
Then \myref{rcpc} is, explicitly,
\beq
Z_N^{(1)} &:& \Phi_{\mu,\nu,i;\mu',\nu',i'} \to
e^{2 \pi i (r_1 + \mu - \mu')/N} \Phi_{\mu,\nu,i;\mu',\nu',i'},
\nnn
Z_N^{(2)} &:& \Phi_{\mu,\nu,i;\mu',\nu',i'} \to
e^{2 \pi i (r_2 + \nu - \nu')/N} \Phi_{\mu,\nu,i;\mu',\nu',i'}.
\eeq
Thus the fields that are retained after the
projection are just those that satisfy
\beq
r_1 + \mu - \mu' = 0 \mod N, \qquad
r_2 + \nu - \nu' = 0 \mod N.
\label{hcon}
\eeq
For instance,
from Table \ref{nalt} we see that components of $x$
must satisfy
\beq
x_{\mu,\nu,i;\mu',\nu',i'} \not= 0 \quad
{\rm iff} \quad \mu'=\mu+1 \mod N, \quad
\nu'=\nu .
\label{xcon}
\eeq

The constraints \myref{hcon} lead to a $N \times N$
periodic lattice interpretation of the surviving components
of $\Phi_{\mu,\nu,i;\mu',\nu',i'}$.  
For instance, \myref{xcon} leaves just components that
can be interpreted as link fields that stretch\footnote{Note
that $\ibf \equiv (1,0)$ and $\jbf \equiv (0,1)$.}
from the site $\mbf \equiv (\mu,\nu)$ 
to $\mbf+\ibf=(\mu+1,\nu)$.  Similarly it
is easy to convince oneself that $y$ yields
link fields stretching from $\mbf \equiv (\mu,\nu)$
to $\mbf+\jbf=(\mu,\nu+1)$.  Note also that $x^\dagger$ and
$y^\dagger$ are link fields oriented in the opposite
direction.  By inspection,
$\alpha$ yields link fermions in the $\ibf$ direction,
$\beta$ link fermions in the $\jbf$ direction, and
$\xi$ link fermions in the $-\ibf-\jbf$ direction.
Since $\mu=\mu'$ and $\nu=\nu'$ for $\lambda$,
it is to be interpreted as a site fermion.
It is convenient to decompose the fields into
$k \times k$ matrices labeled by
site indices of the $N \times N$ lattice:
\beq
x_{\mu,\nu,i;\mu+1,\nu,j} \equiv (x_{(\mu,\nu)})_{i,j} \equiv
(x_\mbf)_{i,j}, \quad {\rm etc.}
\eeq
I summarize these conclusions in Table \ref{uiir}.

\begin{table}
\begin{center}
$$
\begin{array}{c|cccccccc}
\hbox{field} & x_\mbf & y_\mbf & x_\mbf^\dagger & y_\mbf^\dagger &
\alpha_\mbf & \beta_\mbf & \xi_\mbf & \lambda_\mbf \\ \hline
\hbox{orientation} & \ibf & \jbf & -\ibf 
& -\jbf & \ibf & \jbf & -\ibf - \jbf & \hbox{site}
\end{array}
$$
\caption{Classification of components that survive, in
terms of lattice fields.  Fields with negative
orientation point back to site $\mbf$, rather
than starting from $\mbf$.
\label{uiir}}
\end{center}
\end{table}

The $U(k)^{N^2}$ subgroup of the gauge symmetry
\myref{mtgs} that survives is just
\beq
x_\mbf \to u_\mbf^\dagger x_\mbf u_{\mbf+\ibf}, \quad
y_\mbf \to u_\mbf^\dagger y_\mbf u_{\mbf+\jbf}, \quad
{\rm etc.}
\eeq
as dictated by Table \ref{uiir}.
Here, the $u_\mbf$ are $U(k)$ fundamental
representation matrices.
The fields $x_\mbf, y_\mbf$, etc., may written in terms of a
Hermitian basis of $k \times k$ generators
($\mu=0,\ldots,k^2-1$ and $a=1,\ldots,k^2-1$):
\beq
T^\mu \in \left\{ \sqrt{\frac{2}{k}} \obf_k, T^a \right\},
\quad (T^a)^\dagger = T^a, \quad
x_\mbf = x_\mbf^\mu T^\mu, \quad
y_\mbf = y_\mbf^\mu T^\mu, \quad {\rm etc.}
\label{cuur}
\eeq
I define the $T^a$ such that
$\tr(T^\mu T^\nu) = 2 \delta^{\mu \nu}$.
Furthermore I define
\beq
\tr (T^\mu T^\nu T^\rho) = \frac{2\sqrt{2}}{\sqrt{k}}~ t^{\mu \nu \rho},
\label{oomn}
\eeq
and remark that (underlining implies all permutations are to
be taken):
\beq
t^{\underline{\mu \rho 0}} = \delta^{\mu \rho}.
\eeq

Note that only $\lambda$ has components that
survive for which $\mu=\mu'$ and $\nu=\nu'$.
Thus it is the only field that has a nonvanishing
projection onto the diagonal Cartan subalgebra 
$U(1)^{kN^2}$ of $U(kN^2)$:
the diagonal components ($i=i'$) of
these site fields $\lambda_\mbf$ correspond to this
subalgebra of the mother theory.  In particular,
the $U(1)_{diag} = \diag \bigotimes_{i,\mbf} U(1)_{i,\mbf}$ 
projection survives:
\beq
\lambda^\alpha T^\alpha \ni \lambda^0 T^0 
= \frac{1}{\sqrt{kN^2}} \lambda^0 \obf_{kN^2},
\qquad \lambda^0 = \frac{\sqtw}{N} \sum_\mbf \lambda_\mbf^0.
\eeq
From the commutator form of the fermion
action in \myref{22mta}, the $\lambda^0$ component
is an ever-present zeromode; i.e., it vanishes identically
from the action and the integral $\int d \lambda^0$
would cause the partition function to vanish 
if were included.  Since it survives the projection,
we are guaranteed to obtain a vanishing fermion
determinant, just as in the mother theory.  This
$U(1)_{diag}$ sector is completely decoupled,
so there is really no reason to have included it
in the first place.  

To avoid this zeromode, one could just as well have
started with the $SU(kN^2)$ mother theory.  The only
price one pays is that the constraint $\tr \lambda=0$
needs to be imposed.  This is just to say
$\sum_{\mu,\nu,i} \lambda_{\mu,\nu,i;\mu,\nu,i} = 0$,
or equivalently in terms of the site fermions,
\beq
\sum_\mbf \tr \lambda_\mbf 
= \sqrt{2k} \sum_\mbf \lambda_\mbf^0 = 0,
\label{projl}
\eeq
where the trace is over the $U(k)$ index.
This nonlocal constraint (w.r.t.~the lattice) 
leads to a nonlocal
fermion action when the constraint is imposed
by substitution, say, of 
\beq
\lambda^0_{\mbf=(N,N)} = -\sum_{\mbf \not= (N,N)}
\lambda^0_\mbf.
\eeq

In App.~\ref{dfm}, I discuss an alternative method
that factors out the zero eigenvalue.  That factorization
method leaves the fermion action local at the price of
having to work in a limit process.  For numerical applications,
I have found that the cost of extrapolating to
the limit is negligible.

After all the projections are taken into account,
one obtains the following {\it daughter theory} action:
\beq
S &=& \frac{1}{g^2} \sum_\mbf \tr \left\{
\half ( \xdag_{\mbf-\ibf} x_{\mbf-\ibf} - x_\mbf \xdag_\mbf
+ \ydag_{\mbf-\jbf} y_{\mbf-\jbf} - y_\mbf \ydag_\mbf)^2
\right. \nnn
&& + \; 2 (\ydag_{\mbf+\ibf} \xdag_\mbf - \xdag_{\mbf+\jbf} \ydag_\mbf)
(x_\mbf y_{\mbf+\ibf} - y_\mbf x_{\mbf+\jbf}) \nnn
&& + \; \sqtw (\alpha_\mbf \xdag_\mbf \lambda_\mbf
- \alpha_{\mbf} \lambda_{\mbf+\ibf} \xdag_{\mbf}) 
+ \sqtw (\beta_\mbf \ydag_\mbf \lambda_\mbf
- \beta_{\mbf} \lambda_{\mbf+\jbf} \ydag_{\mbf}) \nnn
&& \left. + \; \sqtw (\alpha_{\mbf+\jbf} \xi_\mbf y_\mbf
- \alpha_\mbf y_{\mbf+\ibf} \xi_{\mbf}) 
+ \sqtw (\beta_\mbf x_{\mbf+\jbf} \xi_\mbf
- \beta_{\mbf+\ibf} \xi_\mbf x_\mbf) \right\} . ~~
\label{dtac}
\eeq

Taking advantage of the decomposition \myref{cuur},
one can organize the fermionic part of action into
the following matrix form:
\beq
S_F = \frac{4}{g^2 \sqrt{k}}
(\alpha_\mbf^\mu, \beta_\mbf^\mu) \cdot M_{\mbf,\nbf}^{\mu \rho}
\cdot \binom{\lambda_\nbf^\rho}{\xi_\nbf^\rho } ~ .
\label{sfdf}
\eeq
The fermion matrix $M_{\mbf,\nbf}^{\mu \rho}$
is given by (sum over $\nu$ implied in the entries):
\beq
M_{\mbf,\nbf}^{\mu \rho} = \(
\begin{array}{c|c}
t_{\mbf, \nbf}^{\mu \nu \rho} \xb_\mbf^\nu
- t_{\mbf, \nbf-\ibf}^{\mu \rho \nu} \xb_\mbf^\nu &
- t_{\mbf, \nbf}^{\mu \nu \rho} y_{\mbf+\ibf}^\nu
+ t_{\mbf, \nbf+\jbf}^{\mu \rho \nu} y_\nbf^\nu \\ \hline
t_{\mbf, \nbf}^{\mu \nu \rho} \yb_\mbf^\nu
- t_{\mbf, \nbf-\jbf}^{\mu \rho \nu} \yb_\mbf^\nu &
t_{\mbf, \nbf}^{\mu \nu \rho} x_{\mbf+\jbf}^\nu
- t_{\mbf, \nbf+\ibf}^{\mu \rho \nu} x_\nbf^\nu
\end{array}
\) ~ .
\label{kjtw}
\eeq
Here I have introduced the compact notation
\beq
t_{\mbf, \nbf}^{\mu \nu \rho} = \delta_{\mbf,\nbf} t^{\mu \nu \rho},
\label{tdfd}
\eeq
where $t^{\mu \nu \rho}$ was defined in \myref{oomn}.

The symmetries that survive have been discussed
in \cite{Cohen:2003xe}.  
The symmetries of the lattice theory forbid non-irrelevant operators that
would break supersymmetry in the continuum limit.
Here I only briefly
summarize them.  On has of course the lattice translation
symmetry and $U(k)$ lattice gauge symmetry that
are apparent from the daughter theory action.
The diagonal $U(1)^3$ subgroup of \myref{22rs}
clearly commutes with the orbifold action
and survives.  There is a nontrivial $Z_2$ point
group that reflects the lattice about a diagonal
link.  On the fields it has the action
\beq
&& x_\mbf \to y_{\mbf'}, \quad x^\dagger_\mbf \to y^\dagger_{\mbf'},
\quad y_\mbf \to x_{\mbf'}, \quad y^\dagger_\mbf \to x^\dagger_{\mbf'}, \nnn
&& \alpha_\mbf \to \beta_{\mbf'}, \quad \beta_\mbf \to \alpha_{\mbf'}, \quad
\lambda_\mbf \to \lambda_{\mbf'}, \quad \xi_\mbf \to -\xi_{\mbf'},
\eeq
where $\mbf=(m_1,m_2)$ and $\mbf'=(m_2,m_1)$.
Lastly, there is an exact remnant of the mother theory
supersymmetry.  Suppressing site indices,
\beq
&& \delta x = -\sqtw i \alpha \eta, \quad
\delta y = -\sqtw i \beta \eta, \quad
\delta \lambda = -i([x^\dagger,x]+[y^\dagger,y])\eta, \nnn
&& \delta \xi = 2i[x^\dagger,y^\dagger]\eta, \quad
\delta x^\dagger= \delta y^\dagger = \delta \alpha = \delta \beta = 0.
\eeq
As usual, $\eta$ is an infinitesmal Grassmann parameter.
It is perhaps disturbing that, e.g., $x$ and $x^\dagger$
transform differently.  One can relabel $x^\dagger$
as $\xb$ and treat these formally as independent
complex matrices, as CKKU have done.  This is not
necessary, however; what matters is that an independent
variation of the matrix and its hermitian conjugate---i.e.,
a variation that violates this conjugation constraint---is
a symmetry of the partition function.  This suffices
to derive associated WT identities for variables
that do satisfy the conjugation constraint.

\subsection{Classical approach to the continuum theory}
\label{cat2}
To obtain a discrete approximation to the
continuum theory, one expands about what I will call
the {\it $a$-configuration:}
\beq
x_\mbf = \frac{1}{a \sqtw} \obf, \qquad
y_\mbf = \frac{1}{a \sqtw} \obf, \qquad \forall \mbf,
\label{ckcf2}
\eeq
keeping $g_2=ga$ and $L=Na$ fixed, treating $a$ as small.
(It is easy to see that $S_0=0$ for this configuration.)
That is, we associate $a$ with a lattice spacing, 
even though it arises originally from
a specific background field configuration.
In this case, one finds that the classical continuum limit 
is nothing but \myref{22ck}, as has been described
in detail by CKKU in \cite{Cohen:2003xe}.

Here I merely summarize the map to the continuum action \myref{22ck}.
At each site we define:\footnote{It has been shown
how to exploit polar decomposition to render gauge
variables compact, and to simplify the symmetry
transformations of the fields associated with
the continuum limit \cite{Unsal:2005yh}.  For brevity, I do
not include these details here.}
\beq
x = \frac{1}{\sqtw} (a^{-1} + s_1 + iv_1), \quad
y = \frac{1}{\sqtw} (a^{-1} + s_2 + iv_2).
\label{xycd}
\eeq
The map to continue variables is then given by:
\beq
&& s(\nbf a) = \frac{1}{\sqtw}( s_{1,\nbf} + i s_{2,\nbf}), \quad
v_m(\nbf a) = (v_{1,\nbf}, v_{2,\nbf}), \nnn
&& \psi(\nbf a) = (\lambda_\nbf, \xi_\nbf)^T, \quad
\psib(\nbf a) = i(\alpha_\nbf, \beta_\nbf).
\label{cf22}
\eeq
The Dirac matrices are:
\beq
\gamma_1=\s_3, \quad \gamma_2=\s_1, \quad \gamma_3=\s_2.
\eeq
The last is the chirality matrix, and projections are
defined with
\beq
P_L = \half(1 - \gamma_3), \quad P_R = \half(1+ \gamma_3).
\eeq
As usual, on replaces
\beq
\sum_\nbf (\cdots) \to \int d^2x (\cdots) .
\eeq
Also, shifted indices that appear in the lattice
action are formally expanded in a Taylor
series expansion of \myref{cf22}.  E.g.,
\beq
x_{\mbf+\ibf} \equiv x((\mbf+\ibf)a) = \sum_{n=0}^\infty 
\frac{a^n}{n!} \p_1^n x(\mbf a).
\label{xexp}
\eeq
The $a \to 0$ limit of the lattice action,
once these replacements have been made, defines
the classical continuum limit.

\subsection{Stabilization of the $a$-configuration}
\label{dynlat}
The classical lattice action contains many zero-action
configurations.  As described above, 
CKKU expand the classical lattice action 
about a particular class of zero-action configurations
that are characterized by a parameter $a$, the $a$-configurations.  
In the limit $a \to 0, \; N \to \infty$,
the classical action tends to \myref{22cont}.
Thus $a$ is interpreted as a lattice spacing.  
However, it is dynamical as it has to
do with a particular configuration for the lattice
fields.
This strategy is based on the ideas of {\it deconstruction}
\cite{Arkani-Hamed:2001ca,Hill:2000mu}.

The validity of the semi-classical expansion
about an $a$-configuration rests on the assumption
that it gives a good approximation to the behavior of the 
full lattice theory.  But all $a$-configurations are
energetically equivalent.  Furthermore,
there exist other zero action configurations that do
not fall into the $a$-configuration class (shown below).  
For these reasons, CKKU deform the action by adding an $a$-dependent 
potential that favors the $a$-configuration.  

CKKU suggest the following
deformation of the bosonic action in an effort to stabilize
the theory near the $a$-configuration \myref{ckcf2}:
\beq
S_B &=& S_0 + \SSB, 
\label{sbt} \\
\SSB &=& \frac{a^2 \mu^2}{2 g^2} \sum_\nbf \tr 
\[ \(x_\nbf \xdag_\nbf -\frac{1}{2a^2}\)^2
+ \(y_\nbf \ydag_\nbf-\frac{1}{2a^2}\)^2 \] .
\label{def22}
\eeq
Here the strength of the deformation is determined
by the quantity $\mu$, which has mass dimension 1.
It is clear that the configuration \myref{ckcf2}
minimizes $\SSB$.
The ``continuum
limit'' then includes sending $a \to 0$ in this potential.

Unfortunately, the deformation $\SSB$ breaks 
the exact supersymmetry of the daughter theory
action (hence the subscript ``SB'' = Symmetry Breaking).  
For this reason CKKU demand that
the strength of $\SSB$ relative to $S_0$, conveyed by $\mu^2$,
be scaled to zero in the thermodynamic
limit ($Na \to \infty$):  
\beq
\mu = 1/cL, \quad c = {\rm fixed}.
\label{thli}
\eeq
Thus we are interested in the effects of
the deformation $\SSB$ subject to this scaling.

In this circumstance, what becomes of
all the pleasing symmetry properties of the daughter theory?
The import of this question rests on the
fact that it is the symmetry group of
the daughter theory action that guarantees the
correct continuum limit.  Without those
symmetries, the effective long distance
theory will be more generic, and not what
we intend.

According to CKKU, the deformation $\SSB$
is rendered harmless by scaling its relative strength
to zero in the thermodynamic limit.  For this to hold true,
it is important that the fluctuations about
the $a$-configuration be small and weakly coupled.
Support for the conclusion that the
fluctuations are sufficiently small has been
provided by the arguments offered by CKKU
in \S{6} of \cite{Cohen:2003xe}.

In much of what follows I will specialize to the case
of $U(2)$.  This is merely because it is the simplest
case and the most efficient to simulate.  In this special
case, $x_\mbf, y_\mbf$ will be unconstrained $2 \times 2$
complex matrices.

\subsection{Renormalization}
I will now sketch the simple argument that has
been used by CKKU to demonstrate that no non-irrelevant operators 
that break the continuum symmetries can be generated in
perturbation theory.  Afterward, I will examine
the merits of the argument in more detail.

Consider the form of the perturbative Wilsonian effective
action, $S_{eff} = S + \Delta S$.
The correction
$\Delta S$ will be a sum of monomial operators $\Ocal_{p,\alpha}$
that respect the exact lattice symmetries.  
Here, $[\Ocal_{p,\alpha}]=p$, and $\alpha$ labels
different p-dimensional operators.  Quite generally,
\beq
\Delta S = \sum_\alpha \frac{1}{g_2^2} \int d^2x d\theta 
~ C_{p,\alpha} \Ocal_{p,\alpha}.
\eeq
Here, $\int d\theta$ is the superspace integration
associated with the $Q^2=0$ subalgebra that is
preserved on the lattice.
Note that $[g_2]=1$ and $[d \theta] = 1/2$.  Hence
on dimensional grounds $[C_{p,\alpha}]=(7/2)-p$.
Suppose that $g_2$ and $a$ are the only
dimensionful quantities in the bare lattice
action.  (I will return to the additional
parameters $\mu$ and $L$ below.)  Since, by assumption, the 
coefficients $C_{p,\alpha}$ have been obtained
from a purely perturbative computation, we can
express them as a power series:
\beq
C_{p,\alpha} = a^{p-\frac{7}{2}} \sum_{\ell} 
c_{p,\alpha;\ell} ~ (g_2 a)^{2\ell},
\label{noLc}
\eeq
where $c_{p,\alpha;\ell}$ are dimensionless coefficients
that are independent of $g_2 a$.  Thus the
counterterm corresponding to $\Ocal_{p,\alpha}$
at the $\ell$th order in perturbation theory
has UV degree $D= -p+\frac{7}{2}-2\ell$.
Operators that will survive in the
continuum limit thus satisfy
\beq
p \leq \frac{7}{2} - 2 \ell,
\label{ndco}
\eeq
corresponding to counterterms with UV degree $D \geq 0$.
On the other hand, in perturbation theory $\Ocal_{p,\alpha}$
will consist of monomials of the elementary fields
and spacetime derivatives of them.  It follows that
$p \geq 0$.  Hence \myref{ndco} can only be satisfied
for $\ell=0,1$.  
Recalling that $\ell$ serves as a loop-counting
parameter, we see that only at 1-loop is it possible 
to generate counterterms with UV degree $D \geq 0$.
This is nothing but the usual result:  a theory
for which all couplings have positive mass dimensions will
be super-renormalizable.

CKKU then show that the only operators that
satisfy this requirement are just those that
already appear in the target theory action
and the constant operator, equivalent to
vacuum energy.  The latter can be removed
by a suitable normal-ordering prescription.
Thus one sees that in perturbation theory the
target theory action is obtained without
fine-tuning.

Now I include the parameters $\mu$ and $L$
in the argument.  Since $\mu= c/L$ with $c$ fixed, there is
only one new dimensionful parameter, which I
take to be $L$.  Then \myref{noLc} generalizes
to:
\beq
C_{p,\alpha} = a^{p-\frac{7}{2}} \sum_{\ell,m} 
c_{p,\alpha;\ell,m} \( \frac{L}{a} \)^m (g_2 a)^{2\ell}.
\label{cwL}
\eeq
Marginal and relevant operators thus satisfy
\beq
p \leq \frac{7}{2} - 2 \ell + m.
\eeq
It is assumed in the CKKU arguments that one
can safely take $L \to \infty$.  
One must believe that there are no $m > 0$ terms.
What is the justification for this?

In the $a \to 0$ limit, the lattice action differs
from the continuum action only by $\ord{a}$ terms;
for instance, the higher order derivatives that appear in
\myref{xexp}.  Let us rescale the boson fields so
that they are dimensionless: $v_m \to g_2 v_m, s \to
g_2 s$.  Taking into account the form of the lattice action,
it is easy to see that on dimensional grounds 
the extra terms are $\ord{a \p_m}$ and $\ord{a g_2}$.
The latter are already accounted for in the
last factor of \myref{cwL}, and do not
give rise to the $L/a$ dependence.
The $\ord{a \p_m}$ terms only make a significant contribution
in the UV domain of integration, and should
not affect the IR behavior.  It follows that
no new IR divergences are introduced by
the irrelevant $\ord{a}$ suppressed operators.\footnote{That
this should be true was suggested to me by Joshua Elliott.}
For these reasons I expect that the
IR regulator that is used to define the continuum
theory will also protect the $L \to 0$ limit of
expressions such as \myref{cwL}; the renormalizations
that are sensitive to this IR cutoff are just
those that already appear in the continuum theory.

A more heuristic way to put the argument is as
follows.  The modifications from the lattice action
all involve irrelevant operators.  These are
only important in the UV, based on renormalization
group arguments.  It follows that they will not
give rise to new IR divergences.  This is not
rigorous and I feel that it deserves a more careful
examination that is beyond the scope of this work.

\subsection{The fermion determinant}
\label{tfd}
In this section I will discuss important properties
of the fermion determinant.  Upon examination of \myref{kjtw},
one suspects that $S_F$ is not hermitian.  Thus
one would expect that the fermion determinant will
have a complex phase that depends on the boson background.
This suspicion was confirmed in \cite{Giedt:2003ve},
based on numerical studies.  The phase of the determinant is 
complex and background dependent, as I now review.

Let $\det \hat M$ be the determinant of the
fermion matrix with the
ever-present zeromode removed.  
This can be obtained through the projection \myref{projl}
described above---equivalent to having started
with an $SU(kN^2)$ mother theory; or, it
can be obtained from a limiting process on
a deformation of $M$, as described
in App.~\ref{dfm}.
Alternatively, one can just compute the
eigenvalues of $M$ and compute the
product of those that are nonzero.
The latter two methods were used in
the numerical studies performed in 
\cite{Giedt:2003ve}.
Then the quantity that needs to be
studied in relation to a complex phase
or sign problem is $\arg \det \hat M$,
the phase of the product of (generically) nonzero eigenvalues.

For the $U(2)$ case I take
\beq
T^\mu \in \{ \obf_2, \s^a \}, \qquad
x_\mbf = x_\mbf^0 + x_\mbf^a \s^a, \qquad
y_\mbf = y_\mbf^0 + y_\mbf^a \s^a .
\label{qqre}
\eeq
Then the fermion matrix is given in \myref{kjtw}, with
\beq
t^{000}=1, \qquad t^{\underline{a00}} = 0, \qquad
t^{\underline{ab0}} = \delta^{ab}, \qquad t^{abc} = i\e^{abc} .
\eeq
The lattice theory is obtained by expansion about
the $a$-configuration:
\beq
x_\mbf^0 = \frac{1}{a\sqtw} + \cdots, \qquad
y_\mbf^0 = \frac{1}{a\sqtw} + \cdots .
\eeq
In the study of $\det \hat M$ that was conducted in \cite{Giedt:2003ve},
I scanned over a Gaussian boson distribution
where $x_\mbf^0,y_\mbf^0$ had a nonzero mean $1/a\sqtw \equiv 1$.
The remainder of the bosons were drawn with mean
zero.  All bosons were taken from distributions
with unit variance.  Though this is not the
same distribution as would be generated dynamically
in a Monte Carlo simulation,
it does give a preliminary survey of the behavior
of the phase of the fermion determinant.

\begin{figure}
\begin{center}
\includegraphics[height=5.0in,width=3.0in,angle=90]{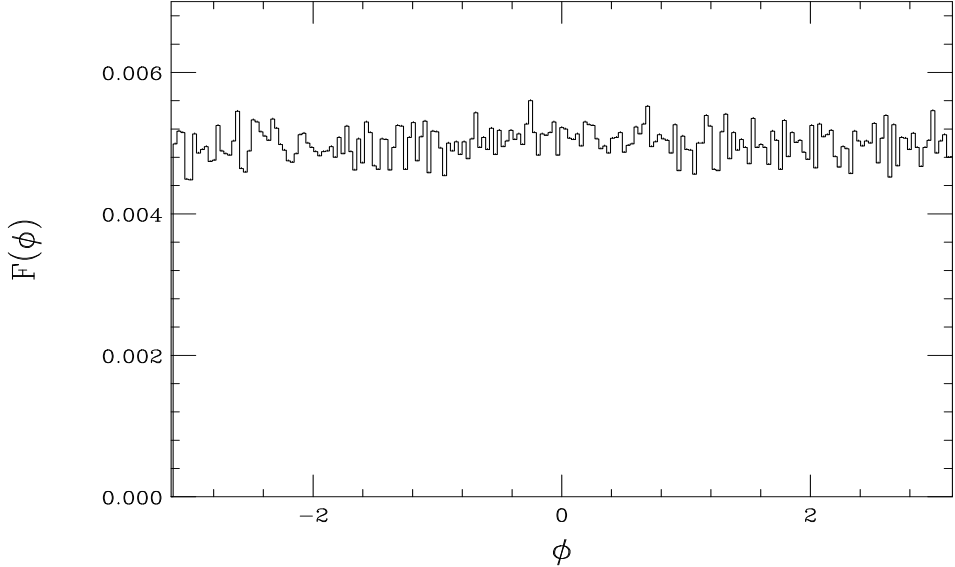}
\end{center}
\caption{Frequency distribution $F(\phi)$
for $\phi = \arg \det \hat M$,
for $10^5$ random (Gaussian) draws, binned into intervals of $\pi/100$.
The distribution of $\phi$ is seen to be nearly 
uniform.  These results are for
the $U(2)$ lattice theory, with $2 \times 2$ lattice.}
\label{f3}
\end{figure}

For a set of $10^5$ draws on the bosons of this
$2 \times 2$ lattice, I 
binned $\arg \det \hat M$ over its range, with
bins of size $\pi/100$.  In Fig.~\ref{f3},
I show the frequency for each bin, as a fraction of
the total number of draws.
The product of the nonzero eigenvalues has arbitrary phase.
Consequently, there is potentially a very serious
problem with cancellations in any Monte Carlo
simulation of this lattice system.  On general
grounds, the cancellations will require an
exponentially increasing number of samples to get
accurate expectation values as the system
volume is increased.  

\subsection{Comparison to $U(k)$ and $SU(k)$ matrix models}

It has already been mentioned that the nonorbifolded
supersymmetric $U(k)$ matrix models---a mother 
theory if $k=k' N^2$---contain ever-present
zeromode fermions coming from the diagonal $U(1)$, as
can be seen from \myref{22mta}.
Then $\det M_{U(k)} \equiv 0$.  However,
it is easy enough to just work with the nonorbifolded $SU(k)$ matrix
model, so that $\psi^0,\psib^0$ are never in the theory to
begin with.  Then with appropriate conventions $\det M_{SU(k)}
\ge 0$.  A proof of this result has been given in
an appendix of \cite{Ambjorn:2000bf}.

One might wonder how the generic phases of Fig.~\ref{f3} are
obtained in the orbifolded matrix models, given the positivity
of the fermion determinant in the nonorbifolded $SU(k)$ matrix models.
That is:  If the fermion determinant of the
mother theory is positive semi-definite,
how can the fermion determinant of the daughter
theory have non-definite complex phase?

Firstly, one should note that the proof given in
\cite{Ambjorn:2000bf} shows only that for each
eigenvalue $\mathfrak{e}$ of $M_{SU(k)}$ there also exists
an eigenvalue ${\mathfrak{e}}^*$, and that these
always come in pairs.  The proof relies essentially on
the relation
\beq
\s^2 v^\alpha \s^2 = \s^2 (v_0^\alpha + i {\bf v}^\alpha
{\bf \cdot} \sbf) \s^2 = (v^\alpha)^* .
\label{jrr}
\eeq
On the other hand, using \myref{v2mat},
and \myref{rcpc} as it applies to $x,y$,
it can be shown that the 
$Z_N \times Z_N$ orbifold action
on the mother theory bosons is generated by
\beq
v^\alpha (T^\alpha)_{i \rho \nu, j \rho' \nu'} &\to&
e^{-i\pi \s^3/N} v^\alpha e^{-i\pi \s^3/N} ~
\Omega_{\rho \mu} (T^\alpha)_{i \mu \nu,j \mu' \nu'} \Omega^{-1}_{\mu' \rho'},
\nnn
v^\alpha (T^\alpha)_{i \rho \mu, j \rho' \mu'} &\to&
e^{i\pi \s^3/N} v^\alpha e^{-i\pi \s^3/N} ~
\Omega_{\rho \nu} (T^\alpha)_{i \mu \nu, j \mu' \nu'} \Omega^{-1}_{\nu' \rho'},
\eeq
where $\Omega$ was defined in \myref{omdf}.
Thus because of the $\exp(\pm i\pi \s^3/N)$ factors,
the orbifold projection does not commute with the
operations of the proof given in \cite{Ambjorn:2000bf};
i.e., Eq.~\myref{jrr}.
The fermion matrix of the 
orbifolded matrix model lacks many
of the eigenvalues of the mother theory; it should come
as no surprise that not all eigenvalues are removed in
pairs $(\mathfrak{e},{\mathfrak{e}}^*)$.  After all,
we already know that only some of the zero eigenvalues
are removed from the $U(kN^2)$ mother theory.
For this reason it is not at all contradictory that the
projected theory has a product of nonzero eigenvalues
that is not positive, nor real.

\subsection{Source of the complex phase problem}
\label{scpp}
I now examine the reasons why the
fermion determinant is not real.  In this discussion,
it is important to note that real-time hermiticity 
takes on a generalized sense in the imaginary-time, 
Euclidean theory.
What is needed is the unitary involution $\Theta$,
satisfying $\Theta^2=1$, that is the continuation
of the real-time hermiticity.  Background for this
discussion is provided in App.~\ref{scap}.

First consider the Euclidean continuum theory \myref{22ck}.
Because of the continuation to imaginary-time (Euclidean)
from real-time (Minkowski), $t_E = -it_M$, the hermiticity
of the real-time action is replaced by
a conjugation that takes $t \to -t$ in the imaginary-time
formulation.  Furthermore, the form of the Dirac adjoint, 
$\psib = \psi^\dagger \gamma_0
\equiv -i \psi^\dagger \gamma_2$, requires that the conjugation
of $\psib,\psi$ in Euclidean space involve factors of $i \gamma_2$.
Finally, the identification $v_2 \equiv iv_0$ requires that the
conjugation change of the sign of $v_2$,
since $v_0$ was taken to be hermitian in real-time.  

It is easily checked that the operation of hermitian
conjugation for the real-time action continues to
the following generalized conjugation in Euclidean space:
\beq
&& \Theta \psi(x,t) = i \psib(x,-t) \gamma_2, \quad
\Theta \psib(x,t) = i \gamma_2 \psi(x,-t), \nnn
&& \Theta v_1(x,t) = v_1(x,-t), \quad
\Theta v_2(x,t) = -v_2(x,-t), \nnn
&& \Theta s(x,t) = s^\dagger(x,-t), \quad
\Theta s^\dagger(x,t) = s(x,-t).
\label{econ}
\eeq
The operator $\Theta$ also conjugates complex numbers
and transposes matrices and expressions involving
fermions.  E.g., $\Theta ( \psib M \psi) = \Theta \psi M^\dagger
\Theta \psib$ for a constant matrix $M$.  It is easily checked that $\Theta^2=1$.
It is a simple exercise to verify that the Euclidean
action \myref{22ck} is self-conjugate under $\Theta$.
That is, $\Theta S = S$.  This must, of course, be
true since the real-time action is hermitian and
$\Theta$ is just a straightforward continuation
of this to imaginary-time.

Now I translate this conjugation to the lattice.
I will choose to associate the $\jbf$ direction
with imaginary-time $x_2=t$, and the $\ibf$ direction
with space $x_1=x$.  Then comparing \myref{econ}
to \myref{xycd}-\myref{cf22}, one sees that the simplest map of
the continuum conjugation\footnote{We could imagine
generalizations that smear nearby sites in a unitary
fashion.} is just:
\beq
&& \Theta x_\mbf = x_{\mbf'}^\dagger, \quad 
\Theta x^\dagger_\mbf = x_{\mbf'}, \quad
\Theta y_\mbf = y_{\mbf'}, \nnn
&& \Theta y_\mbf^\dagger = y_{\mbf'}^\dagger, \quad
\Theta \lambda_\mbf = -\beta_{\mbf'}, \quad
\Theta \xi_\mbf = - \alpha_{\mbf'}, \nnn
&& \Theta \alpha_\mbf = -\xi_{\mbf'}, \quad
\Theta \beta_\mbf = -\lambda_{\mbf'}, \nnn
&& {\rm with} \; \mbf' \equiv (m_1,-m_2) 
\quad {\rm where} \; \mbf=(m_1,m_2).
\label{thla}
\eeq
That $y$ does not transform into $y^\dagger$ just
follows from the $v_2 \to -v_2$ property of \myref{econ}
and the complex conjugation property of $\Theta$ when
acting on complex numbers.

It is straightforward to check that the
lattice action $S_T=S+S_{SB}$, given by \myref{dtac}
and \myref{def22}, is not self-conjugate
with respect to \myref{thla}.  On the other
hand, as has already been mentioned, we can
express the approach to the continuum limit
in terms of a power series of continuum expressions:
\beq
S_T = S^{(0)} + aS^{(1)} + a^2 S^{(2)} + \cdots .
\eeq
Here, $S^{(0)}$ is just the continuum action \myref{22ck}.
As has been stated, $\Theta S^{(0)} = S^{(0)}$.  It
follows that
\beq
\Theta S_T - S_T = \Theta ( aS^{(1)} + a^2 S^{(2)} + \cdots )
= \ord{a}.
\eeq
Thus, the violation of the continuum self-conjugacy is
entirely due to the irrelevant, $\ord{a}$ suppressed
operators.

In particular, it is not difficult to show that
\beq
\Theta S_F - S_F &=& \frac{a \sqtw}{g^2} \sum_\mbf \tr \bigg\{
- \alpha_\mbf \nabla_2^- x^\dagger_\mbf \lambda_\mbf
- \beta_{\mbf-\jbf} \lambda_\mbf \nabla_2^- y_\mbf^\dagger \nnn
&& + \alpha_{\mbf+\jbf} \xi_\mbf \nabla_2^+ y_\mbf
+ \beta_m \nabla_2^+ x_\mbf \xi_\mbf \bigg\}.
\label{sfvi}
\eeq
Here, forward and backward difference operators
are used:
\beq
a \nabla_2^+ x_\mbf = x_{\mbf+\jbf} - x_\mbf, \quad
a \nabla_2^- x_\mbf = x_\mbf - x_{\mbf-\jbf}.
\eeq
From \myref{sfvi} one sees that terms that
violate self-conjugacy of the fermion action
are $\ord{pa}$ when we pass to momentum space.
These will not be small for the modes with
$p = \ord{a^{-1}}$.  The fermion determinant
includes the product of eigenvalues associated
with these modes, independent of how small
we take the lattice spacing $a$.  It is therefore
not surprising that we do not recover a real
fermion determinant.  We always have eigenvalues
that reflect the lack of self-conjugacy of the
fermion action.

Lastly, I remark that the boson lattice action \myref{sbt}
is not self-conjugate w.r.t.~to \myref{thla} either.
The violation is, according to the general arguments
made above, $\ord{a}$.  On the other hand, the
boson lattice action is self-conjugate under
ordinary hermitian conjugation.  On the $a$-configuration,
the fermion lattice action reduces to
\beq
S_{F0} = \frac{1}{g^2} \sum_\mbf \tr \left\{
(\nabla_1^- \alpha_\mbf + \nabla_2^- \beta_\mbf) \lambda_\mbf
+ (\nabla_2^+ \alpha_\mbf - \nabla_1^+ \beta_\mbf) \xi_\mbf \right\}.
\eeq
Then if we define
\beq
\lambda_\mbf^\dagger = \lambda_\mbf, \quad
\xi_\mbf^\dagger = \xi_\mbf, \quad
\alpha_\mbf^\dagger = -\alpha_\mbf, \quad
\beta_\mbf^\dagger = -\beta_\mbf,
\eeq
one finds that $S_{F0}^\dagger = S_{F0}$.  This property
is easily extended to the non-irrelevant interaction
terms involving the vector-boson and scalars.  On the
other hand, it can be shown that the irrrelevant, $\ord{a}$
suppressed terms do not enjoy this property.  This
too is consistent with the fact that the
fermion determinant is complex.


\mys{Deconstructed lattice (4,4) 2d SYM}
\label{ql44}
As a further example, I now consider another 2d SYM
theory.  
In \cite{Cohen:2003qw}, CKKU proposed a lattice action 
for (4,4) 2d SYM.  In \cite{Giedt:2003vy}, some aspects
of the fermion determinant were examined, with results
very similar to those for the (2,2) model.  I will
briefly present those results below. 
I will more extensively review my efforts \cite{Giedt:2004tn} to study
a rather fundamental aspect of the construction:
the proposed emergence of the type of dynamical lattice 
spacing that was described for the
(2,2) model above.  The semi-classical approach
lends much support to the assertion that
the continuum approach is exactly as CKKU have
proposed.  It would be nice to obtain this
same result from Monte Carlo similulations of
the model, which I have attempted.  Unfortunately,
the complex phase problem has left me with only
supportive but inconclusive results, as will be seen.

\subsection{The continuum theory}

One begins with 6d YM minimally coupled to single massless 
symplectic Majorana fermion in the adjoint representation.
This is $\Ncal=1$ 6d SYM; then (4,4) 2d SYM is obtained as
its dimensional reduction to 2d.
Similar to the reduction described above,
$g_2^2 = g_6^2/R^4$, and again $[g_2]=1$.
Note that the 4 vector boson modes along the
compactified dimensions are considered as real scalars
from the 2d point of view.
As usual, the notation ``(4,4)'' denotes that in the 2d
theory there are 4 left-moving, left-chirality
supercurrents, and 4 right-moving,
right-chirality supercurrents.

Continuing to a Euclidean formulation,
one finds that the 2d Euclidean action is
\beq
S &=& \int d^2 x \; \frac{1}{g_2^2} \tr \[ (Ds_\mu) \cdot (Ds_\mu)
+ \psib_i \Dslash \psi_i + \fourth F \cdot F \right. \nnn
&& \left. 
+ \psib_i [ (s_0 \delta_{ij} + i \gamma_3 {\bf s} \cdot \sbf_{ij}) ,\psi_j] 
- \half [s_\mu, s_\nu]^2 \],
\label{wkkr}
\eeq
where $s_\mu$ ($\mu=0,1,2,3$) are hermitian scalars, $F$ is
the 2d YM field strength, and $\psi_i$ ($i=1,2$) are
2d Majorana fermions, all in the adjoint representation.
For our purposes, the gauge group will be $U(k)$.

\subsection{The mother and daughter theories}
The mother theory is obtained from the Euclidean
$\Ncal=1$ 6d $U(kN^2)$ super-Yang-Mills.
The action is dimensionally 
reduced to 0d to obtain a $U(kN^2)$ matrix model.
The matrix model naturally possesses $SO(6) \simeq SU(4)$ 
Euclidean invariance.
Next we note $SU(4) \supset U(1) \times U(1) \supset 
Z_N \times Z_N$.  This is very much like $SO(4) \simeq
SU(2) \times SU(2) \supset U(1) \times U(1) \supset Z_N \times
Z_N$ of the (2,2) model above.
CKKU identify a homomorphic embedding 
of $Z_N \times Z_N$ into the $U(kN^2)$ gauge symmetry group,
just as in Eqs.~\myref{omdf}-\myref{c12df} of the (2,2) theory.
Using this, a $Z_N \times Z_N$ orbifold projection is performed to
obtain a $U(k)^{N^2}$ 0d {\it quiver} (or {\it moose}) theory.
Due to the similarity to the (2,2) model,
I will not review the details
of this procedure, but just state the end result.

The bosonic part of the daughter theory action is
\beq
S_0 &=& \frac{1}{g^2} \tr \sum_{\nbf} \[
\half (\xdag_{\nbf-\ibf} x_{\nbf-\ibf} - x_{\nbf} \xdag_{\nbf}
+ \ydag_{\nbf-\jbf} y_{\nbf-\jbf} - y_\nbf \ydag_\nbf
+ \zdag_\nbf z_\nbf - z_\nbf \zdag_\nbf )^2 \right. \nnn
&& + 2(x_\nbf y_{\nbf+\ibf} - y_\nbf x_{\nbf+\jbf})
(\ydag_{\nbf+\ibf} \xdag_\nbf - \xdag_{\nbf+\jbf} \ydag_\nbf) \nnn
&& + 2(y_\nbf z_{\nbf+\jbf} - z_\nbf y_\nbf)
(\zdag_{\nbf+\jbf} \ydag_\nbf - \ydag_\nbf \zdag_\nbf) \nnn 
&& \left. + 2(z_\nbf x_\nbf - x_\nbf z_{\nbf+\ibf})
(\xdag_\nbf \zdag_\nbf - \zdag_{\nbf+\ibf} \xdag_\nbf) \]
\label{sbta}
\eeq
Here, $x_\mbf, y_\mbf, z_\mbf$ are bosonic lattice
fields that are $k \times k$ unconstrained
complex matrices.
The $U(k)^{N^2}$ symmetry is nothing but the local $U(k)$
symmetry of the lattice action $S_0$, with link bosons $x_\mbf$ in
the $\ibf$ direction, link bosons $y_\mbf$ in the $\jbf$ direction, and
sites bosons $z_\mbf$, all transforming in the usual manner:
\beq
x_\mbf \to \alpha_\mbf x_\mbf \alpha_{\mbf + \ibf}^\dagger, \qquad
y_\mbf \to \alpha_\mbf y_\mbf \alpha_{\mbf + \jbf}^\dagger, \qquad
z_\mbf \to \alpha_\mbf z_\mbf \alpha^\dagger_\mbf .
\label{yuer}
\eeq
Canonical mass dimension 1 is assigned to $x_\mbf, y_\mbf, z_\mbf$,
whereas $g$ has mass dimension~2.

The fermionic lattice action can be written in the form

\beq
S_F = - \frac{1}{g^2}
\( \psi_{1,\mbf}^\mu ~,~ \psi_{2,\mbf}^\mu ~,~
\psi_{3,\mbf}^\mu ~,~ \chi_\mbf^\mu \)
\cdot M_{\mbf \nbf}^{\mu \rho} \cdot
\begin{pmatrix} \xi_{1, \nbf}^\rho \cr
\xi_{2, \nbf}^\rho \cr \xi_{3, \nbf}^\rho
\cr \lambda_\nbf^\rho \cr \end{pmatrix} 
\label{haht}
\eeq
The elements of the fermion matrix are:
\beq
(M_{\mbf \nbf}^{\mu \rho})_{1,1} &=&
(M_{\mbf \nbf}^{\mu \rho})_{2,2} \; = \;
(M_{\mbf \nbf}^{\mu \rho})_{3,3} \; = \;
(M_{\mbf \nbf}^{\mu \rho})_{4,4} \; = \; 0,
\nnn
(M_{\mbf \nbf}^{\mu \rho})_{1,2} &=&
- \tmr_{\mbf,\nbf} z_{\nbf+\ibf}^\nu + \tmn_{\mbf,\nbf} z_\nbf^\nu ,
\quad
(M_{\mbf \nbf}^{\mu \rho})_{1,3} =
\tmr_{\mbf,\nbf} y_{\nbf+\ibf}^\nu - \tmn_{\mbf,\nbf+\jbf} y_\nbf^\nu ,
\nnn
(M_{\mbf \nbf}^{\mu \rho})_{1,4} &=&
\tmr_{\mbf,\nbf} \xb_{\nbf}^\nu - \tmn_{\mbf,\nbf-\ibf} \xb_\mbf^\nu ,
\quad
(M_{\mbf \nbf}^{\mu \rho})_{2,1} =
\tmr_{\mbf,\nbf} z_{\nbf+\jbf}^\nu - \tmn_{\mbf,\nbf} z_\nbf^\nu ,
\nnn
(M_{\mbf \nbf}^{\mu \rho})_{2,3} &=&
- \tmr_{\mbf,\nbf} x_{\nbf+\jbf}^\nu + \tmn_{\mbf,\nbf+\ibf} x_\nbf^\nu ,
\quad
(M_{\mbf \nbf}^{\mu \rho})_{2,4} =
\tmr_{\mbf,\nbf} \yb_{\nbf}^\nu - \tmn_{\mbf,\nbf-\jbf} \yb_\mbf^\nu ,
\nnn
(M_{\mbf \nbf}^{\mu \rho})_{3,1} &=&
- \tmr_{\mbf,\nbf} y_{\nbf}^\nu + \tmn_{\mbf,\nbf+\jbf} y_\nbf^\nu ,
\quad
(M_{\mbf \nbf}^{\mu \rho})_{3,2} =
\tmr_{\mbf,\nbf} x_{\nbf}^\nu - \tmn_{\mbf,\nbf+\ibf} x_\nbf^\nu ,
\nnn
(M_{\mbf \nbf}^{\mu \rho})_{3,4} &=&
\tmr_{\mbf,\nbf} \zb_{\nbf}^\nu - \tmn_{\mbf,\nbf} \zb_\nbf^\nu ,
\quad
(M_{\mbf \nbf}^{\mu \rho})_{4,1} =
- \tmr_{\mbf,\nbf} \xb_{\nbf+\jbf}^\nu
+ \tmn_{\mbf,\nbf-\ibf} \xb_\mbf^\nu ,
\nnn
(M_{\mbf \nbf}^{\mu \rho})_{4,2} &=&
- \tmr_{\mbf,\nbf} \yb_{\nbf+\ibf}^\nu
+ \tmn_{\mbf,\nbf-\jbf} \yb_\mbf^\nu ,
\quad
(M_{\mbf \nbf}^{\mu \rho})_{4,3} =
- \tmr_{\mbf,\nbf} \zb_{\nbf+\ibf+\jbf}^\nu
+ \tmn_{\mbf,\nbf} \zb_\nbf^\nu . ~ ~
\label{hahu}
\eeq
The notation is as in the (2,2) model above.
Note in particular the definitions \myref{oomn} and \myref{tdfd}.

This matrix has two ever-present fermion zeromodes.
The first is associated with left multiplication
(the transpose ``T'' merely indicates a row vector,
for consistency with \myref{haht}):
\beq
(\xi_{1, \nbf}^\rho , \xi_{2, \nbf}^\rho , \xi_{3, \nbf}^\rho ,
\lambda_\nbf^\rho )^T
= (0 , 0 , 0 , \lambda \delta^{\rho 0} )^T \quad
\forall \quad \nbf .
\label{zv1}
\eeq
The second is associated with right multiplication:
\beq
\( \psi_{1,\mbf}^\mu ~,~ \psi_{2,\mbf}^\mu ~,~
\psi_{3,\mbf}^\mu ~,~ \chi_\mbf^\mu \)
= \( 0 ~,~ 0 ~,~
\psi \delta^{0 \mu} ~,~ 0 \) \quad \forall \quad \mbf
\label{zv2}
\eeq
The eigenvalues of $M$ are defined through left
multiplication.  Only \myref{zv1} corresponds to
an ever-present eigenvalue 0 of $M$, whereas
\myref{zv2} corresponds to an ever-present
eigenvalue 0 of $M^T$.
Because the matrix $M$ is not hermitian,
it is diagonalized by $M \to D = U M V$
with $U$ and $V$ independent unitary matrices.
When this is done, the diagonal matrix $D$
has just 1 zero on the diagonal, corresponding
to a column of $V$ and a row of $U$, from
\myref{zv1} and \myref{zv2}.  These ever-present
zeromodes originate from site fermions
that are associated with the $U(1)_{{\rm diag}}$
subgroup of the mother theory.
Having identified these modes in the daughter
theory, it is straightforward to project them
out as in \myref{projl}.

\subsection{Construction of the lattice theory}
\label{clth44}
To obtain the lattice theory, one
proceeds in analogy to the (2,2) theory.
An expansion is made about the $a$-configuration
\beq
x_\mbf = \frac{1}{a \sqtw} \obf, \quad
y_\mbf = \frac{1}{a \sqtw} \obf, \quad
z_\mbf = 0, \quad \forall \mbf,
\label{ckcf}
\eeq
keeping $g_2=ga$ and $L=Na$ fixed, treating $a$ as small.
It is easy to see that $S_0=0$ for this configuration.
As before, we associate $a$ with a lattice spacing, 
even though it arises from
a background field configuration.
One finds that the classical continuum limit 
is \myref{wkkr}, as has been shown in detail
in \cite{Cohen:2003qw}.

In order to make the configuration \myref{ckcf}
energetically preferred, CKKU utilize
a deformation that is just a slight
modification of the one for the (2,2) theory, Eq.~\myref{def22}:
\beq
S_B &=& S_0 + \SSB, \\
\SSB &=& \frac{a^2 \mu^2}{2 g^2} \sum_\nbf \tr 
\[ \(x_\nbf \xdag_\nbf -\frac{1}{2a^2}\)^2
+ \(y_\nbf \ydag_\nbf-\frac{1}{2a^2}\)^2 
+ \frac{2}{a^2} z_\nbf \zdag_\nbf \] .
\eeq
That is, a term is added to stabilize the
$z$-field about zero.
As in the (2,2) case, $\SSB$ breaks the exact supersymmetry of 
the daughter theory action.  Once again CKKU demand that
the strength of $\SSB$ relative to $S_0$, conveyed by $\mu^2$,
be scaled to zero in the thermodynamic limit \myref{thli}.

As in the (2,2) case, I will generally specialize to $U(2)$ 
gauge theory in what follows.

\subsection{The semi-classical analysis}
\label{s:cla}
As will be explained below, the bosonic
action of the undeformed daughter theory
satisfies $S_0 \geq 0$ and a vast number of nontrivial solutions 
to $S_0=0$ exist, not all of which are gauge equivalent.  
In fact, the space of minimum
action configurations, or {\it moduli space,} is a multi-dimensional
noncompact manifold with various {\it branches}---classes of
configurations.  In \S\ref{jirr},
some of these minima will be illustrated.  In
\S\ref{wqfe} I will consider the modifications induced
by the deformation $\SSB$, which has the effect of
lifting all flat directions in moduli space,
modulo gauge equivalences.
That is, the $a$-configuration is the
absolute minimum of $S_B = S_0 + S_{SB}$.
However, because the deformation is scaled to
zero in the thermodynamic limit, it is important
to keep in mind the undeformed moduli space.
Through understanding this classical picture,
naive expections of what will occur in the quantum
theory, based on energetics, can be formulated.
In \S\ref{dms}, the effect of the
fermion determinant $\det \hat M$ will be
considered.  It will be shown that the minima of the
undeformed bosonic action $S_{0}$ also have $\det \hat M=0$.
Significantly, this is also true of the $a$-configuration.
Fluctuations about these minima of $S_{0}$ cause
the determinant to be nonvanishing; however, for
small fluctuations I find that the magnitude
of the determinant differs significantly depending
on which minimum they perturb.
This gives some hints as to the 
relative contribution of these various
configurations.  My conclusion will be that
the continuum limit is the target
theory, but that a renormalization
of the dynamical lattice spacing occurs.

\subsubsection{Undeformed theory}
\label{jirr}
Here I neglect $\SSB$ and examine the minima of $S_0$.
Note that \myref{sbta} is a sum of terms of the form
$\tr A A^\dagger$ (the first line involves squares of
hermitian matrices).  Thus $S_0 \geq 0$ with $S_0 = 0$
iff the following equations hold true:
\beq
&& \xdag_{\nbf-\ibf} x_{\nbf-\ibf} - x_{\nbf} \xdag_{\nbf}
+ \ydag_{\nbf-\jbf} y_{\nbf-\jbf} - y_\nbf \ydag_\nbf
+ [\zdag_\nbf, z_\nbf] = 0 ,
\label{wrse} \\
&& x_\nbf y_{\nbf+\ibf} - y_\nbf x_{\nbf+\jbf} = 
y_\nbf z_{\nbf+\jbf} - z_\nbf y_\nbf = 
z_\nbf x_\nbf - x_\nbf z_{\nbf+\ibf} = 0 .
\label{wrsf}
\eeq
The set
of solutions is the classical moduli space of the undeformed theory.

\bfe{Zeromode branch.}
To begin a study of the moduli space,
I isolate the zero momentum modes:  $x_\nbf \equiv x \; \forall
\nbf$, etc.  Then Eqs.~\myref{wrse} and \myref{wrsf} reduce to
\beq
&& [\xdag,x] + [\ydag,y] + [\zdag,z] = 0, \nnn
&& [x,y] = [y,z] = [z,x] = 0.
\label{qers}
\eeq
Eqs.~\myref{qers}
may be recognized as nothing but the {\it D-flatness}
and {\it F-flatness} constraints that describe
the moduli space associated with the classical
vacuum of $\Ncal=4$ 4d super-Yang-Mills.\footnote{I
thank Erich Poppitz for pointing this out to me,
as well as the branch of moduli space \myref{kerr} given below.}
The equations are invariant with respect to the global
gauge transformation
\beq
x \to \alpha x \alpha^\dagger, \quad
y \to \alpha y \alpha^\dagger, \quad
z \to \alpha z \alpha^\dagger.
\label{oore}
\eeq
Then it is well-known that solutions to \myref{qers}
consist of $x,y,z$ that lie in a Cartan subalgebra
of $U(k)$; the proof is reviewed in App.~\ref{n4ms}.  
The global gauge transformations \myref{oore}
allow one to change to a basis where this Cartan subalgebra
has a diagonal realization.  Thus one can think of the
moduli space as the set of all possible diagonal matrices $x,y,z$,
and all global gauge transformations 
\myref{oore} of this set.\footnote{Actually,
the gauge transformations \myref{yuer} 
that follow from \myref{wrse}-\myref{wrsf}
also leave the equations \myref{qers} invariant.  This
generates an inhomogeneous solution to $S_0=0$, and
will be important in \S\ref{prov} below.}

In particular, the zeromode moduli space of the undeformed
$U(2)$ theory is completely described by
\beq
x = x^0 + x^3 \s^3, \quad 
y = y^0 + y^3 \s^3, \quad 
z = z^0 + z^3 \s^3,
\label{uier}
\eeq
with arbitrary complex numbers $x^0,x^3,y^0,y^3,z^0,z^3$,
together with $U(2)$ transformations of these solutions.

Eqs.~\myref{wrse} and \myref{wrsf} also have non-zeromode
solutions.  I do not attempt to present an exhaustive
account of them.  I will merely point out a few such branches
in order to illustrate that the undeformed theory
has a very complicated and large set of $S_0=0$ configurations.

\bfe{$x_\mbf=y_\mbf=0$ branch.}
This is the branch of moduli space described by
\beq
x_\mbf=y_\mbf=0, \quad z_\mbf = z_\mbf^0 + z_\mbf^3 \s^3, \quad
\forall \mbf.
\label{kerr}
\eeq
Again, $z_\mbf^0,z_\mbf^3$ are arbitrary complex numbers.
Furthermore, $z_\mbf$ is a site variable and thus transforms
independently at each site as
\beq
z_\mbf \to \alpha_\mbf z_\mbf \alpha^\dagger_\mbf.
\eeq

It can be seen that this branch affords a vast number of solutions to
\myref{wrse} and \myref{wrsf}; there are $N^2$ such
solutions, modulo choices for $z_\mbf^0, z_\mbf^3 \in \Cbf$
and gauge equivalences.

\bfe{$z_\nbf =0$ branch.}
Another branch in moduli space is the following.  First I
set $z_\nbf =0 \; \forall \nbf$, and introduce Fourier space
variables
\beq
x_\nbf = \frac{1}{N} \sum_\kbf \omega^{\kbf \cdot \nbf} f_\kbf,
\qquad
y_\nbf = \frac{1}{N} \sum_\kbf \omega^{\kbf \cdot \nbf} g_\kbf,
\qquad
\omega = \exp(2\pi i/N),
\eeq
where $\kbf=(k_1,k_2)$ and $k_1, k_2 \in [0,1,\ldots,N-1]$.
Then taking into account $z_\nbf=0$, the conditions \myref{wrse}
and \myref{wrsf} are equivalent to:
\beq
0 &=& \sum_\kbf \( \omega^{\ibf \cdot \lbf} f_\kbf^\dagger f_{\kbf-\lbf}
- f_\kbf f_{\kbf+\lbf}^\dagger
+ \omega^{\jbf \cdot \lbf} g_\kbf^\dagger g_{\kbf-\lbf}
- g_\kbf g_{\kbf+\lbf}^\dagger \), \nnn
0 &=& \sum_\kbf \( \omega^{-\ibf \cdot (\lbf + \kbf)} f_\kbf g_{-\kbf-\lbf}
- \omega^{\jbf \cdot \kbf} g_{-\kbf-\lbf} f_\kbf \),
\label{klsf}
\eeq
for all $\lbf=(\ell_1,\ell_2)$ and $\ell_1, \ell_2 \in [0,1,\ldots,N-1]$.
Next I turn off all modes except one for both $f_\kbf$ and $g_\kbf$:
\beq
f_\kbf = \delta_{\kbf,\kbf'} f_{\kbf'}, \qquad
g_\kbf = \delta_{\kbf,-\kbf'} g_{-\kbf'}.
\eeq
Here and below, no sum over $\kbf'$ is implied.
When substituted into \myref{klsf}, only 2 nontrivial
conditions survive:
\beq
0 = [f_{\kbf'}^\dagger, f_{\kbf'}]
+ [g_{-\kbf'}^\dagger, g_{-\kbf'}], \qquad
0 = f_{\kbf'} g_{-\kbf'} 
- \omega^{(\ibf+\jbf) \cdot \kbf'} g_{-\kbf'} f_{\kbf'}.
\label{klsg}
\eeq
For the $U(2)$ case, solutions exist if
$\omega^{(\ibf+\jbf)\cdot \kbf'} = \pm 1$.  

For $\omega^{(\ibf+\jbf)\cdot \kbf'} = 1$
we have solutions for $f_{\kbf'}, g_{-\kbf'}$
diagonal matrices.  This includes the zeromode
solution ($\kbf'=0$) \myref{uier} with
$z \equiv 0$.  The $a$-configuration \myref{ckcf}
is a special case.  I want to emphasize that
any statement that holds generally for the $z_\nbf=0$
branch will hold in particular for the $a$-configuration.

In the case of $\omega^{(\ibf+\jbf)\cdot \kbf'} = -1$
it is easy to see that there are solutions of the form
\beq
f_{\kbf'} = z_f \s^3, \quad g_{-\kbf'} = z_g (\s^1 + b \s^2),
\quad z_f, z_g \in \Cbf, \quad b \in \Rbf.
\label{ppor}
\eeq
There are many values of $\kbf'$ for
which $\omega^{(\ibf+\jbf)\cdot \kbf'} = \pm 1$.  
For $N$ even, these are
\beq
k'_1 + k'_2 = 0, \frac{N}{2}, N, \frac{3N}{2}.
\eeq
For $N$ odd, $k'_1 + k'_2 = 0, N$ are allowed and in the cases
where
\beq
k'_1 + k'_2 = \frac{N \pm 1}{2}, \frac{3 (N\pm 1)}{2}
\label{ppot}
\eeq
\myref{ppor} yield approximate solutions to \myref{klsg}, with an error
of order $1/N$.  

Thus in the $N \to \infty$ limit the number 
of $S_0=0$ configurations in the $z_\nbf=0$ class is vast; in fact, it
is easy to check that the number of such configurations is
approximately $2N$, modulo gauge equivalences and
various choices for the constants such as in \myref{ppor}.

\subsubsection{Deformed theory}
\label{wqfe}
Now I consider the supersymmetry breaking
deformation $\SSB$ introduced by CKKU.  
To see its effect it is handy to rewrite
the quantities that appear in it.  Recall that $x_\mbf$ is a
complex $2 \times 2$ matrix.  Dropping the subscript, we
can always define
\beq
x=x^0 + x^a \s^a, \qquad x^\dagger = \xb^0 + \xb^a \s^a.
\eeq
Then it is straightforward to work out ($\mu=0,\ldots,3$)
\beq
x x^\dagger = x^\mu \xb^\mu + (x^0 \xb^c + \xb^0 x^c
+ i x^a \xb^b \e^{abc} ) \s^c \equiv \phi^{x,0} + \phi^{x,c} \s^c
\equiv \phi^x.
\label{xxdd}
\eeq
Note that $\phi^{x,\mu}$ are real, and
that $\phi^{x,0}$ is positive definite.  With
similar definitions for $\phi^y, \phi^z$, the CKKU deformation is
\beq
\SSB &=& \frac{a^2 \mu^2}{2 g^2} \sum_\mbf \tr \[ \(\phi_\mbf^x-\frac{1}{2a^2}\)^2
+ \(\phi_\mbf^y-\frac{1}{2a^2}\)^2 + \frac{2}{a^2} \phi_\mbf^z \] \nnn
&=& \frac{a^2 \mu^2}{g^2} \sum_\mbf \[ \( \phi_\mbf^{x,0} - \frac{1}{2a^2}\)^2
+ \( \phi_\mbf^{y,0} - \frac{1}{2a^2}\)^2 
+ \frac{2}{a^2} \phi_\mbf^{z,0} \right. \nnn 
&& \quad \left. + \sum_a \[ (\phi_\mbf^{x,a})^2 + (\phi_\mbf^{y,a})^2 \] \].
\eeq
It can be seen that the deformation drives $\phi_\mbf^{x,a},\phi_\mbf^{y,a},
\phi_\mbf^{z,0}$ toward the origin, and $\phi_\mbf^{x,0}, \phi_\mbf^{y,0}$ 
toward $1/2a^2$.  When $\phi_\mbf^{z,0}=0$, 
it is easy to see that $\phi_\mbf^{z,a}=0$ identically.

To continue the analysis, 
it is convenient to rescale to dimensionless quantities---denoted
by a ``hat''---using
the parameter $a$:
\beq
\hat g = g a^2, \quad \hat \mu = \mu a, 
\quad \hat \phi_\mbf^x = a^2 \phi_\mbf^x,
\quad \hx_\mbf = a x_\mbf, \quad {\rm etc.}
\label{krkr}
\eeq
Then $\SSB =$
\beq
\frac{{\hat \mu}^2}{{\hat g}^2} 
\sum_\mbf \[ \( \hphi_\mbf^{x,0} - \frac{1}{2}\)^2
+ \( \hphi_\mbf^{y,0} - \frac{1}{2}\)^2 
+ 2 \hphi_\mbf^{z,0}
+ \sum_a \[ (\hphi_\mbf^{x,a})^2 
+ (\hphi_\mbf^{y,a})^2 \] \].
\label{pool}
\eeq
For any value of the lattice spacing $a$, the minimum of $\SSB$ is
obtained iff
\beq
\hphi_\mbf^{x,0} = \hphi_\mbf^{y,0}= \half, \quad
\hphi_\mbf^{z,0} = \hphi_\mbf^{x,a} =\hphi_\mbf^{y,a} = 0,
\quad \forall \mbf.
\label{jjus}
\eeq
For example,
from \myref{xxdd} one sees that the 
conditions involving $\hx_\mbf$ are just
\beq
\hx_\mbf^\mu \hxb_\mbf^\mu = \half, \qquad
\hx_\mbf^0 \hxb_\mbf^c + \hxb_\mbf^0 \hx_\mbf^c 
+ i \hx_\mbf^a \hxb_\mbf^b \e^{abc} = 0.
\label{piwr}
\eeq
Let us examine what additional constraint this places
on classical solutions to $S=0$, beyond the restrictions
\myref{wrse}-\myref{wrsf} of the undeformed theory.  

First I note that most of the non-zeromode configurations
discussed in \S\ref{jirr} above are not minima of
$\SSB$.  Non-zeromode cases that are 
minima will be discussed in \S\ref{prov} below.
For the zeromode configurations \myref{uier},
Eqs.~\myref{piwr} imply
that \myref{uier} is restricted to the form
\beq
\hx = \frac{e^{i \gamma_x}}{\sqtw} \diag \( e^{i \varphi_x},e^{-i \varphi_x} \)
\label{xsom}
\eeq
and symmetry transformations of this.
That is, $\hx$ is restricted to be an element of the 
maximal abelian subgroup $U(1)^2$ of $U(2)$, up to an
overall factor of $1/\sqtw$.  Similarly, we have for $\hy$,
\beq
\hy = \frac{e^{i \gamma_y}}{\sqtw} 
\diag \( e^{i \varphi_y},e^{-i \varphi_y} \).
\label{ysom}
\eeq
Finally, $\hphi_\mbf^{z,0} = 0$ implies $z_\mbf^\mu \zb_\mbf^\mu = 0$,
which has the unique solution $z_\mbf=0, \; \forall \mbf$.

There is a $U(1)^4$ global symmetry of the daughter
theory, described in \cite{Cohen:2003qw}, 
that can be used to remove the phases $\gamma_x,\gamma_y$.
Apart from global obstructions that are essentially Polyakov loops
in the $\ibf$ or $\jbf$ directions, the phases $\varphi_x,\varphi_y$ in
the configuration \myref{xsom}
and \myref{ysom} can be gauged away.
It is straightforward to
verify that the required gauge transformation is \myref{yuer} with
\beq
\alpha_{m_1,m_2} = \diag \( e^{i(m_1 \varphi_x + m_2 \varphi_y)},
e^{- i(m_1 \varphi_x + m_2 \varphi_y)} \).
\eeq
This sets all $\hx_\mbf,\hy_\mbf$ to $1/\sqtw$
except at the boundaries---the location of
which is a matter of
convention due to translation invariance and
periodic boundary conditions:
\beq
\hx_{N,m_2} &=& \frac{1}{\sqtw} 
\diag \( e^{i N \varphi_x},e^{-i N\varphi_x} \), \quad \forall m_2; \nnn
\hy_{m_1,N} &=& \frac{1}{\sqtw} 
\diag \( e^{i N \varphi_y},e^{-i N\varphi_y} \), \quad \forall m_1.
\eeq
For most purposes, I do not expect such vacua to distinguish
themselves from the trivial vacua in the thermodynamic limit.
In any case, global features such as these are typical of classical
vacua of other lattice Yang-Mills formulations, such as the Wilson
action.  In my simulation study (cf.~\S\ref{sima}), I have avoided this issue by
restricting my attention to the expectation value of
a quantity that is independent of these angles.

\subsubsection{Inhomogeneous minima of the deformed theory}
\label{prov}
Consider the $z_\mbf=0$ branch with $\omega^{(\ibf+\jbf)\cdot \kbf}
= 1$, where I have dropped the prime on $\kbf'$.  
Following the details given above, one has
\beq
x_\mbf = \frac{1}{N} \omega^{\kbf \cdot \mbf} f_{\kbf}, \quad
y_\mbf = \frac{1}{N} \omega^{-\kbf \cdot \mbf} g_{-\kbf},
\label{zmzz}
\eeq
with $f_{\kbf},g_{-\kbf}$ diagonal matrices.  That is,
\beq
f_{\kbf} = b + d \s^3, \quad g_{-\kbf} = b' + d' \s^3, \quad
b,d,b',d' \in \Cbf.
\eeq
In the notation of \myref{xxdd} and \myref{krkr}, it is easy to check that
\beq
\hphi^{x,0} = \frac{a^2}{N^2} ( |b|^2 + |d|^2), \quad
\hphi^{x,c} = \frac{a^2}{N^2} (b \bar d + \bar b d) \delta^{c3},
\label{hpsi}
\eeq
and similar expressions for $\hphi^y$.  From \myref{pool}
one sees that a minimum of $S_{SB}$ is obtained provided
\beq
b \bar d + \bar b d = b' \bar d' + \bar b' d' = 0, 
\quad |b|^2 + |d|^2 = |b'|^2 + |d'|^2 = \frac{N^2}{2a^2}.
\label{pore}
\eeq
A simple parameterization of these minima is the following:
\beq
b = \frac{N}{a \sqtw} e^{i \theta} \cos \eta, \quad 
d = i \frac{N}{a \sqtw} e^{i \theta} \sin \eta \quad 
\Rightarrow \quad
b + d \s^3 = \frac{N}{a \sqtw} e^{i \theta} e^{i \eta \s^3},
\eeq
with similar definitions for the primed variables.
Equivalently,
\beq
x_\mbf = \frac{1}{a \sqtw} \omega^{\kbf \cdot \mbf} 
e^{i \theta} e^{i \eta \s^3}, \quad
y_\mbf = \frac{1}{a \sqtw} \omega^{-\kbf \cdot \mbf} 
e^{i \theta'} e^{i \eta' \s^3}.
\label{xynm}
\eeq
The resulting 2d $\times$ 2d space of
minima of the total bosonic action, $S_B = S_0+S_{SB}$,
occurs for each value of $\kbf$ s.t.
\beq
k_2 = -k_1 \mod N.
\label{kcon}
\eeq

Next I note that with \myref{pore} imposed,
\beq
\tr \s^3 x_\mbf x^\dagger_\mbf = \frac{2}{N^2} (b \bar d + \bar b d) = 0.
\eeq
This raises the possibity that \myref{xynm} are
gauge-equivalent to the $a$-configuration, or
more generally the configurations \myref{xsom}-\myref{ysom}
that differ from the $a$-configuration only by edge
effects.  I now show that this is the case.

Since by assumption the
matrices appearing in \myref{zmzz} are diagonal,
we can restrict to the diagonal $U(1)^2$ subgroup
of the $U(2)$ gauge transformations \myref{yuer}.
Furthermore, the $\mbf$ dependence in \myref{zmzz}
is an overall phase, so we can restrict further
to the $U(1)_{{\rm diag}}$ subgroup.  Thus I
take $\alpha_\mbf = \exp (i \zeta_\mbf)$ in \myref{yuer}.
The assumption that \myref{xynm} are gauge
transformations of \myref{xsom}-\myref{ysom} implies:
\beq
&& \theta = \gamma_x, \quad \eta = \varphi_x, \quad
\theta' = \gamma_y, \quad \eta'= \varphi_y, \\
&& e^{i(\zeta_\mbf - \zeta_{\mbf+\ibf})} = \omega^{\kbf \cdot \mbf}, \quad
e^{i(\zeta_\mbf - \zeta_{\mbf+\jbf})} = \omega^{-\kbf \cdot \mbf}.
\label{xyas}
\eeq
It is not hard to solve the recursion relations \myref{xyas}:
\beq
\zeta_{(m_1,m_2)} = \zeta_{(1,1)} - \frac{\pi k_1}{N} \[ (m_1
- m_2)^2 + (2-m_1-m_2) \],
\label{xyat}
\eeq
with the phase $\zeta_{(1,1)}$ at site $(1,1)$ arbitrary.
To prove that these satisfy \myref{xyas}, it is important
to keep \myref{kcon} in mind.

Similarly, $S_{SB}=0$ solutions exist on the $\omega^{\kbf
\cdot (\ibf + \jbf)} = -1$ part of the $z_\mbf=0$ branch.
From \myref{ppor},
\beq
x_\mbf x_\mbf^\dagger = \frac{1}{N^2} |z_f|^2, \quad
y_\mbf y_\mbf^\dagger = \frac{1}{N^2} |z_g|^2 (1 + b^2).
\eeq
Choosing $|z_f|^2 = |z_g|^2 (1 + b^2) = N^2/2a^2$ gives the
minima.  However, these too are gauge equivalent to the
configurations \myref{xsom}-\myref{ysom}.  The gauge
transformations are more complicated due to the
nondiagonal nature of \myref{ppor} and the constraint
\beq
k_2 = -k_1 + \frac{N}{2} \mod N.
\label{kco2}
\eeq
The conditions that must be satisfied by the
gauge transformation parameters \myref{yuer} are:
\beq
\alpha_{\mbf+\ibf} = \omega^{-\kbf \cdot \mbf} \alpha_\mbf
~ (i \s^3), \quad
\alpha_{\mbf+\jbf} = \omega^{\kbf \cdot \mbf} \alpha_\mbf
~ [i( \cos \phi \s^1+ \sin \phi \s^2)]
\eeq
where $\phi = \tan^{-1} b$ with $b$ defined in \myref{ppor}.
These recursion relations are solved by
\beq
\alpha_\mbf &=& e^{i \zeta_\mbf} \alpha_{(1,1)}
[i( \cos \phi \s^1+ \sin \phi \s^2)]^{m_2-1}
(i \s^3)^{m_1-1}, \nnn
\zeta_\mbf &=& \zeta_{(1,1)} + \frac{\pi}{N} [
m_2(m_2-2m_1+1)k_2 \nnn
&& + (2(m_2-1)-m_1(m_1-1))k_1 ].
\eeq

To summarize, I find that with the addition of
the deformation $S_{SB}$, the $a$-configuration
is the only one that persists as a classical minimum,
modulo gauge equivalent configurations and
boundary effects.  The $z_\mbf=0$ branch consists
of the zeromode branch \myref{uier} with $z=0$,
and gauge transformations of this.

However, the strength $\mu$ of the deformation
is scaled to zero in the thermodynamic
limit.  One might wonder if the $x_\mbf=y_\mbf=0$ minima
of $S_0$ that were lifted by $S_{SB}$
become important again in this limit.
I next show that this is not the case, due
to a suppression from the fermion determinant.\footnote{
This is to be contrasted with what happens if the fermions
are not included \cite{Giedt:2003gf}.}

\subsubsection{Determinant on moduli space}
\label{dms}
I now show that for each of the boson configurations discussed
in \S\ref{jirr} above---branches of the undeformed theory moduli
space---additional fermion zeromodes appear.
It will be seen that this is also true for the $a$-configuration.
In configuration space, the weights of
the immediate neighborhoods of these $S_0=0$ points are
damped by fermion effects, since $\det \hat M \approx 0$.
(Recall that in $\hat M$ the ever-present fermion zeromode
has been projected out.)  Of course, even in
the continuum theory the trivial vacuum---equivalent 
to the $a$-configuration---has a
vanishing fermion determinant (for periodic boundary
conditions).  What is important
is the relative weight of the various neighborhoods
of the different minima of $S_0$.  Since the
strength of $S_{SB}$ is scaled to zero in
the thermodynamic limit, considerations
other than the $\exp (- S_{SB})$ suppression
come into play when one attempts to determine the correct
configurations that should be used for a semiclassical expansion.
The determination of the dominant saddlepoint of the
effective action---about
which the continuum limit is best defined---is a 
subtle question; it involves an interplay between the size of
$\det \hat M \times \exp(-S_B)$ and entropic effects.
It will be seen that the $a$-configuration dominates.

\bfe{The $a$-configuration.}
It is a simple matter to work out the
fermion matrix for the $a$-configuration \myref{ckcf}
from \myref{hahu}.  Then one can check that
in \myref{haht}
\beq
(\xi_{1, \nbf}^\rho , \xi_{2, \nbf}^\rho , \xi_{3, \nbf}^\rho ,
\lambda_\nbf^\rho )^T =
\delta^{\rho 0} ( \xi_1, \xi_2, \xi_3, \lambda)^T ~~ \forall \nbf,
\label{aczm}
\eeq
is a zeromode for any values 
of $\xi_i,\lambda$.
It follows from the considerations of App.~\ref{dfm} that
$\det M_\e(v) \equiv \det (M(v)+\e) \propto \e^4$, where $v$ is the
$a$-configuration.  Moreover, if $1 \gg \delta \gg \e$,
we have $\det M_\e(v + \ord{\delta}) \propto \e \delta^3$,
since the $\ord{\e}$ eigenvalues not associated with
ever-present zeromodes are shifted by $\ord{\delta}$.
The conclusion of this is that for order $\delta$ fluctuations 
about the $a$-configuration, the suppression due
to approximate zeromodes is $\det \hat M = \ord{\delta^3}$.
Next I will compare this to the other points
of the $S_0=0$ moduli space that were discussed
in \S\ref{jirr}.

\bfe{$x_\mbf=y_\mbf=0$ branch.}  
Here it is easily checked that $M$ acting on the vector
\beq
( \xi_{1, \nbf}^\rho , \xi_{2, \nbf}^\rho , \xi_{3, \nbf}^\rho ,
\lambda_\nbf^\rho )^T = ( 0 , 0 , 0 , \lambda_\nbf \delta^{\rho 0} )^T
\eeq
vanishes for any choice of the $N^2$ Grassmann variables
$\lambda_\nbf$.  This gives $N^2$ fermion zeromodes associated
with left multiplication.
Thus $\det M_\e(v) \propto \e^{N^2}$,
where $v$ denotes the selected point along the
$x_\mbf=y_\mbf=0$ branch.
Again introducing
fluctuations of order $\delta$, with
$1 \gg \delta \gg \e$, we see that 
$\det M_\e \propto \e \delta^{N^2-1}$,
leading to a $\det \hat M \propto \delta^{N^2-1}$
suppression. Thus the
inclusion of fermions dramatically changes the
weight of configuration space in the neighborhood of
$x_\mbf=y_\mbf=0$ branch in the integration
measure.  Relative to the $a$-configuration,
there is an $\ord{N^2 \delta^{N^2-4}}$ suppression,
where the factor $N^2$ comes from the number
of solutions \myref{kerr}.
In the thermodynamic limit ($N \to \infty$),
this branch has vanishing weight
in comparison to the $a$-configuration.
In a semi-classical analysis of the
continuum limit it need not be included.

\bfe{$z_\mbf=0$ branch.}
Due to the gauge equivalence that was established
in \S\ref{prov}, we already know that the number
of fermion zeromodes for $S_{SB}=0$ configurations
will be the same as for the $a$-configuration.  However,
the configurations with $S_{SB} \not= 0$ also
have the same number of fermion zeromodes.
This is not surprising, since other configurations
on the $z_\mbf=0$ branch correspond to minima of
$S_{SB}$ with some other value for the parameter $a$.

First consider the case where $\omega^{k_1+k_2} = 1$.
Then set
\beq
( \xi_{1, \nbf}^\rho , \xi_{2, \nbf}^\rho , \xi_{3, \nbf}^\rho ,
\lambda_\nbf^\rho )^T = \delta^{\rho 0}
( \omega^{\kbf \cdot \nbf} \xi_1 ,
\omega^{-\kbf \cdot \nbf} \xi_2 , \xi_3 ,
\lambda )^T \quad \forall \quad \nbf ,
\label{zmz01}
\eeq
with $\xi_i,\lambda$ arbitrary Grassmann numbers.
It is easy to check that these are fermion zeromodes on
this branch.  The $\lambda$ direction is just the
ever-present zeromode.  The 3 $\xi_i$ directions are
particular to this branch.  It can be seen that
the $a$-configuration zeromodes \myref{aczm} are a
special case of \myref{zmz01}, corresponding to $\kbf=0$.
Clearly, the supression of fluctuations about
this branch, due to approximate fermion
zermodes, is of the same order for all points
on the branch: $\det M_\e \propto \e \delta^3$, 
and hence $\det \hat M \propto \delta^3$.
It is also simple to see that the fermion zeromodes
are related to those of the $a$-configuration
by the gauge transformation \myref{xyas}-\myref{xyat} that relates
the two configurations, as must be the case.

Now consider the case where $\omega^{k_1+k_2} = -1$.
Then set
\beq
( \xi_{1, \nbf}^\rho , \xi_{2, \nbf}^\rho , \xi_{3, \nbf}^\rho ,
\lambda_\nbf^\rho )^T = \delta^{\rho 0} ( \omega^{-\kbf \cdot \nbf} \xi_1 ,
\omega^{\kbf \cdot \nbf} \xi_2 , (-)^{n_1+n_2} \xi_3 ,
\lambda )^T \quad \forall \quad \nbf .
\eeq
with $\xi_i,\lambda$ arbitrary Grassmann numbers.
It is not hard to check that these are fermion zeromodes on
this branch.  

\bfe{Summary.}  
Due to the gauge equivalence of the $a$-configuration
and other $S_0=S_{SB}=0$ configurations, and due to the suppression
of the $x_\mbf=y_\mbf=0$ branch, semi-classical expansion
about the $a$-configuration suffices to obtain
reliable estimates in the lattice theory.  In particular,
one is interested in the effective lattice spacing,
\beq
\frac{1}{a^2_{{\rm eff}}} \equiv \vev{\tr x_\mbf x_\mbf^\dagger}
= \vev{\tr y_\mbf y_\mbf^\dagger},
\label{efls}
\eeq
in the $a \to 0, N \to \infty$ limit.  A
renormalization of the lattice spacing has already been studied
perturbatively about the $a$-configuration in \cite{Onogi:2005cz}.
Another approach to studying this question is
by Monte Carlo simulations, which I now take up.

\subsection{Simulation analysis of the dynamic lattice spacing}
\label{sima}
In the present subsection I review 
Monte Carlo simulations \cite{Giedt:2004tn}
that investigated
the stability of the $a$-configuration.
Some idea of the effect of the fermion
determinant has already been gained from
the semi-classical analysis of \S\ref{dms}.
The intent of the simulation analysis was to
pursue a more systematic and complete analysis.
I have studied
\beq
\vev{\hphi_\mbf^{x,0}} = \vev{\hx_\mbf^\mu \hxb_\mbf^\mu}
= \bigvev{ \half \tr (\hx_\mbf \hxd_\mbf) } = a^2_{{\rm eff}}/2 a^2
\label{iwre}
\eeq
in my simulations.  This expectation value
is to be compared to the classical 
prediction~\myref{jjus}.

\subsubsection{Scaling}
I study \myref{iwre} along a naive scaling trajectory:
\beq
g_2 = a^{-1} \hat g(a) = \mbox{fixed} .
\label{yuie}
\eeq
That is, I hold the bare coupling in physical units, $g_2$, fixed.
The dimensionless bare coupling $\hat g$ is 
then a function of $a$ that vanishes
linearly with $a$ as the UV cutoff is removed.

With regard to $\hmu$ I follow the instructions of CKKU:  
I send the dimensionless coefficient $\hat \mu$ of the deformation $\SSB$ to
zero as $1/N$ while increasing $N$:
\beq
\hat \mu^{-1} = c N,  \qquad c = \ord{1}
\label{pwer}
\eeq
where $c$ is a constant.  This is equivalent to \myref{thli}.

In the rescaled variables \myref{krkr},
the coefficient of the undeformed action is $1/\hg^2$,
whereas the coefficient of the deformation is $\hmu^2/ \hg^2$,
as can be seen from \myref{pool}.
The relative strength of the deformation vanishes
in the thermodynamic limit, when \myref{pwer} is imposed.

I perform these scalings for a sequence of decreasing
values of $a$.  I then
extrapolate toward $a=0$ to obtain the continuum limit.
The length scales (in lattice units) are set by 
$\hg_2^{-1}$ and the system size $N$.
To keep discretization and finite-size effects to a minimum,
I would like to take $1 \ll \hg^{-1} \ll N$, but
often violate the bounds $1 \leq \hg^{-1} \leq N$ for specific
points where measurements are taken.  The reason for this
is that data outside the optimal window 
is informative to the extrapolation.

\subsubsection{Problems with the complex phase}
It is essential to
include the fermions if we are to draw conclusions for the
supersymmetric system.
The present system suffers from a complex
fermion determinant \cite{Giedt:2003vy}.  The integration
measure $\det \hat M e^{-S_B}$ is not positive semi-definite
and does not provide a satisfactory probability measure
for Monte Carlo simulations.  If we
factor out the phase $e^{i \alpha} = \det \hat M / |\det \hat M|$ 
and use $|\det \hat M| e^{-S_B}$
instead, then we will generate the {\it phase-quenched (p.q.)}
ensemble of boson configurations.  Expectation values
of an operator $\Ocal$ in the full theory are formally
related to those in the phase-quenched theory by the
{\it reweighting} identity
\beq
\vev{\Ocal} = \vev{e^{i \alpha} \Ocal}_{p.q.} /
\vev{e^{i \alpha}}_{p.q.}
\label{tslr}
\eeq

\begin{figure}
\begin{center}
\includegraphics[height=4.5in,width=3.0in,angle=90]{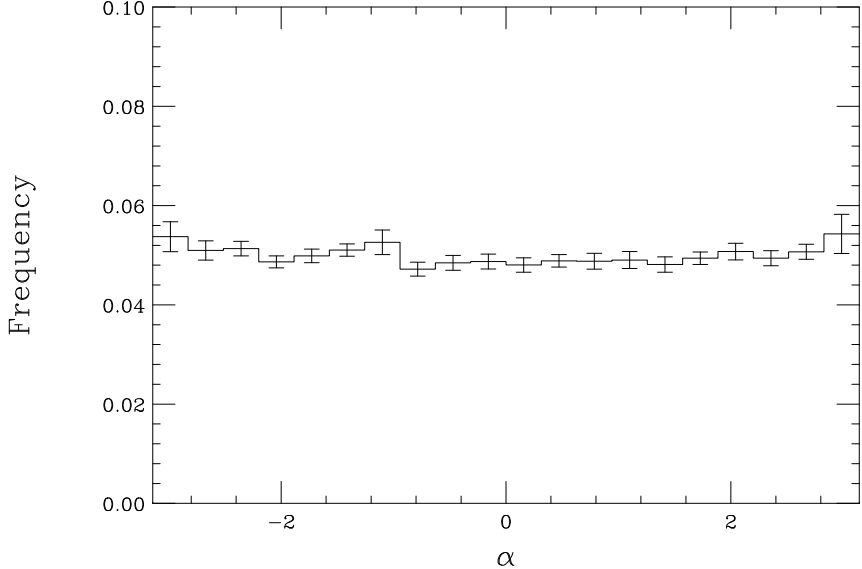}
\end{center}
\caption{Average frequency distribution 
for $\alpha = \arg \det \hat M$ in the phase-quenched distribution
for the $2 \times 2$ lattice.
Error bars are determined by variance in bin counts for several
blocks of data. \label{phdb}}
\end{figure}

In some lattice systems, the distribution of $\alpha$ in the 
phase-quenched ensemble is sharply peaked. 
The phase-quenched quantities in the ratio exist 
and can be measured reliably.  For example, this is the case in the 
4d $U(1)_L \times U(1)_R$ symmetric Yukawa model of \cite{Munster:1992jq}.
By contrast, the results for the
CKKU system are not encouraging.  The distribution of $\alpha
= \arg \det \hat M$ in the phase-quenched ensemble 
was found in \cite{Giedt:2003vy} to be essentially flat,
as can be seen from Fig.~\ref{phdb}.  This leads to approximate
cancellations when I attempt to compute reweighted quantities
contained in the numerator and denominator of \myref{tslr}.
More details will be given below.  

As we saw in Sec.~\ref{dms}, $\det \hat M$ vanishes for
the $S_0=0$ configurations that were discussed above.
The phase-quenched quantity 
$\vev{\hphi_\mbf^{x,0}}_{p.q.}$
takes this suppression into account.  
I therefore believe it gives some hints
as to how the CKKU proposal for a dynamical
lattice spacing works out.  In Sec.~\ref{pooe},
I return to the matter of phase reweighting.

\subsubsection{Results for the phase-quenched ensemble}
\label{oope}
\begin{figure}
\begin{center}
\includegraphics[height=4.5in,width=3in,angle=90]{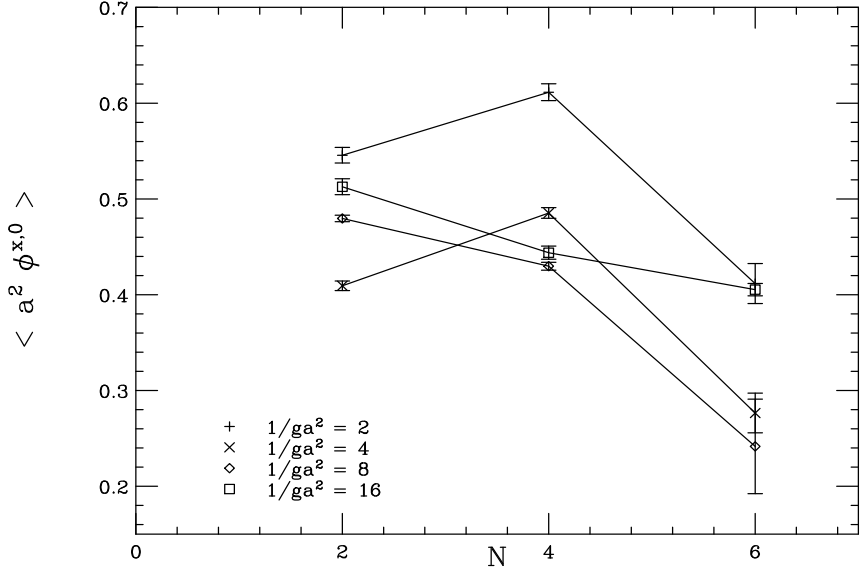}
\end{center}
\caption{Phase-quenched averages (lines to guide the eye). 
\label{lupq}}
\end{figure}

In Fig.~\ref{lupq}, 
$\vev{\hat \phi_\mbf^{x,0}}$ is obtained in the phase-quenched
ensemble; that is, the fermion determinant was taken into
account in Metropolis updates, using the linear variation
\beq
\delta \ln |\det \hat M| = \real \tr [ \hat M^{-1} \delta \hat M ]
\eeq

I remind the reader that larger values of $\hginv = 1/g a^2$
correspond to finer lattices, and thus extrapolate to the
continuum limit.  However, to hold $L=Na$ fixed,
successive trajectories must be compared
as $N$ versus $2N$, since $\hginv$ is doubled.  It can be
seen from Fig.~\ref{lupq} that for $N \leq 6$ it is not 
possible to reliably extrapolate
to the large $L$ behavior.  However, the $g a^2 = 1/16$ results
do look encouraging.

\begin{figure}
\begin{center}
\includegraphics[height=4.5in,width=3in,angle=90]{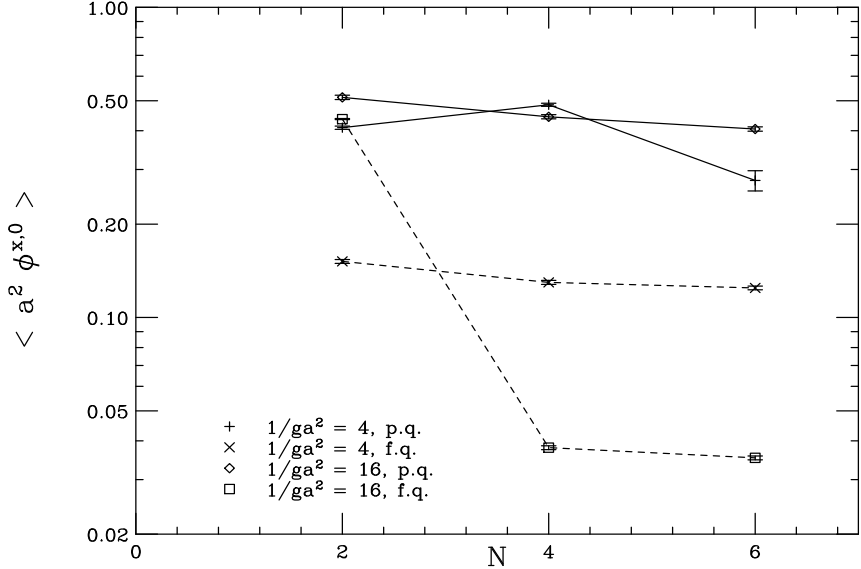}
\end{center}
\caption{A comparison of some of the
averages obtained in phase-quenched (solid lines to
guide the eye) versus
fully-quenched (dashed lines to guide the eye) simulations.
Note that the vertical axis in on a log scale.
\label{lupqfq}}
\end{figure}

In Fig.~\ref{lupqfq} I provide a comparison between the
phase-quenched and fully-quenched results.  It can
be seen that the effect of the fermion determinant
is dramatic.  For the finest lattice that I have
available, $\hginv = 16$, the phase-quenched results are
more than an order of magnitude larger when we look
at $N=6$.  Furthermore, the phase-quenched results are
always of the same order of magnitude as the classical
estimate~\myref{jjus}.

\subsubsection{Phase reweighting}
\label{pooe}
As was reported in \cite{Giedt:2003vy}, and
mentioned above, the distribution
of $\alpha = \arg \det \hat M$ is essentially flat with respect
to the phase-quenched ensemble.  In Fig.~\ref{phdb} I present
results for a $2 \times 2$ lattice.  It can be seen that it
will be quite difficult, if not impossible, to obtain a
reliable estimate of $\vev{\exp(i \alpha)}_{p.q.}$.

E.g., I find that for $N=2$, $\hginv=2$, with 1000 samples,
the ``noise-to-signal'' ratio is
\beq
\s_{\vev{\cos \alpha }_{p.q.}} / \vev{\cos \alpha }_{p.q.} =
\s_{\vev{\sin \alpha }_{p.q.}} / \vev{\sin \alpha }_{p.q.} = \ord{1} .
\eeq
For $N=4$, $\hginv=2$, with 1000 samples, I find
\beq
\s_{\vev{\cos \alpha }_{p.q.}} / \vev{\cos \alpha }_{p.q.} = 
\s_{\vev{\sin \alpha }_{p.q.}} / \vev{\sin \alpha }_{p.q.} = \ord{10} .
\eeq
On general grounds, these noise-to-signal ratios are expected to
get exponentially worse as $N$ is increased.
Similarly discouraging results are found for the other quantities
that need to be estimated in order to make use of the reweighting
identity \myref{tslr}.  Since reweighting does not even work for
small $N$, it looks like a futile approach by which to study
the full theory.

However, the complex phase is not present
in the target continuum theory; so, it should be possible to
address the phase problem.  
As described above in \S\ref{scpp},
simply taking $a \to 0$ is
not sufficient at a nonperturbative level.
The irrelevant operators that violate the self-conjugacy
of the fermion action are no longer
suppressed for modes with momentum $p \sim 1/a$.
For this reason, the phase of the fermion determinant
will always fluctuate wildly.
I am currently studying ways to 
improve on the CKKU construction and eliminate the
complex phase problem based on this new understanding.

\mys{Conclusions}
\label{s:con}
As was explained in \S\ref{moti}, many motivations
for the study of supersymmetric field theories
by lattice techniques exist.  In \S\ref{s:vap} we
saw that many approaches to lattice supersymmetry have
been attempted.  The plethora of formulations is
due to the fact that no one of them represents an entirely
satisfactory general method.

In the remainder of this report I have reviewed
one of the more promising approaches, the one
based on dimensional deconstruction.  Two 2d SYM
theories were described, as examples.  The (2,2)
model described in \S\ref{ql22} exhibits the basic
CKKU technique for construction of the class of
models.  I reviewed the complex phase difficulty 
that occurs for the fermion determinant.  This
problem also arised in the (4,4) model of \S\ref{ql44}.
I described how this phase originates from
irrelevant, $\ord{pa}$ suppressed operators that
are not present in the continuum.  I further
explained that since the fermion determinant
includes modes up to $p=\ord{a^{-1}}$, the
complex phase does not disappear in the $a \to 0$
limit.

For the (4,4) model that was described in \S\ref{ql44},
I studied the classical moduli space of the
undeformed theory.  I then examined the properties
of the fermion determinant on this moduli space.
It was found that this determinant suppresses
one branch of the moduli space---the one that
would have vanishing effective lattice spacing---but
does not suppress another branch---one that would have
nonvanishing effective lattice spacing.  
I showed that the class of
minima of the deformed bosonic action
is gauge-equivalent to the $a$-configuration,
modulo boundary effects.

Attempts to study the effective lattice spacing by Monte Carlo
simulations were briefly described.  It was
seen that the phase-quenched averages---which
include the suppression due to the fermion determinant---support
the basic picture of CKKU:  an effective
lattice spacing survives when the deformation
is scaled to zero in the thermodynamic limit.
Reweighting by the phase proved to be hopeless,
due to a nearly flat distribution.

It is my hope to present further results 
regarding renormalization of the
effective lattice space, using the semiclassical approach,
in a future publication.
I am currently studying formulations that will 
overcome the phase problem, based on the
new understanding of its origin that was
presented in \S\ref{scpp}.

As mentioned in \S\ref{qecs}, I am currently
investigating the Sugino construction of lattice
SYM \cite{Sugino:2003yb,Sugino:2004qd,Sugino:2004uv,Sugino:2006uf}.  
In a forthcoming article I will give details regarding
the lack of self-conjugacy for the fermion action,
due to irrelevant operators, quite similar to
what occurs for the CKKU models.  I also
hope to show how this problem can be alleviated
through improvements to the action.

Lastly, I mentioned in \S\ref{nlqm} that I am
currently investigating the relationship between
nonlocal invariances of the lattice partition
function and symmetries of the continuum theory.
Here the issue is that an invariance that
touches all time slices simultaneously does not
have a simple interpretation as the action of
an operator that commutes with the transfer
matrix, and hence hamiltonian.  What is needed
is to show that the operator associated with the
transformation localizes in the continuum limit.

\vspace{15pt}

\noindent
{\bf \Large Acknowledgements}

\vspace{5pt}

\noindent
I would like to thank Erich Poppitz for discussion
and comments.  Thanks are also due to a referee of an early
version of \cite{Giedt:2003gf}, who provided 
useful criticism and suggestions.  I thank Mithat \"Unsal
for useful communications, and David B.~Kaplan for
helpful discussions.
Early stages of this work were conducted at the University
of Toronto, Canada, with
support from the National Science and Engineering 
Research Council of Canada and a grant provided
through the Ontario Premier's Research Excellence Award.
Later stages of this review were performed at
the University of Minnesota,
with support from the U.S.~Department of Energy
under grant No.~DE-FG02-94ER-40823.

\myappendix

\mys{A toy (2,2) theory}
\label{mnnot}
In this section I clarify aspects of the
``(2,2)'' \susy\ notation, using the minimal (2,2) free
scalar multiplet theory.
There, the Euclidean action takes the form
\beq
S = \int d^2x \; \[ - 4 \phib \p_z \p_\zb \phi 
- \Fb F + 2i \psi_+ \p_z \psib_+ 
- 2i \psib_- \p_\zb \psi_- \] .
\label{cact}
\eeq
Here the subscripts on the fermions correspond
to an axial $U(1)$ that leaves the scalars
neutral: 
\beq
\psi_\pm \to e^{\pm i \alpha} \psi_\pm, \quad
\psib_\pm \to e^{\mp i \alpha} \psib_\pm .
\label{uia}
\eeq

With periodic boundary conditions for all fields,
the action is invariant under the infinitesmal
supersymmetry transformations
\beq
\delta \phi &=& \e^- \psi_- + \e^+ \psi_+ , \qquad
\delta \psi_+ = -2i \eb^+ \p_\zb \phi + \e^- F , \nnn
\delta \psi_- &=&  2i \eb^- \p_z \phi - \e^+ F , \qquad
\delta F = 2i \eb^- \p_z \psi_+ + 2i \eb^+ \p_\zb \psi_-,  \nnn
\delta \phib &=& - \eb^- \psib_- - \eb^+ \psib_+ , \qquad
\delta \psib_+ = 2i \e^+ \p_\zb \phib + \eb^- \Fb, \nnn
\delta \psib_- &=& -2i \e^- \p_z \phib - \eb^+ \Fb , \qquad
\delta \Fb = 2i \e^- \p_z \psib_+ + 2i \e^+ \p_\zb \psib_- .
\eeq
On the other hand, if we make the $\e$'s position dependent,
\beq
\delta S = 4 \int d^2x \[ \p_\zb \e^- S_- + \bar S_- \p_\zb \eb^-
+ \p_z \e^+ S_+ + \bar S_+ \p_z \eb^+ \],
\eeq
where
\beq
S_- = \psi_- \p_z \phib, \quad \bar S_- = \psib_- \p_z \phi,
\quad S_+ = \psi_+ \p_\zb \phib, \quad \bar S_+ = \psib_+ \p_\zb \phi.
\eeq
It can be seen from \myref{uia} that $S_-$, and hence the corresponding
superalgebra, is of left chirality, and that $S_+$ is of right
chirality.

Also note that the equations of motion
\beq
\p_z \psi_+ = \p_z \psib_+ = \p_\zb \psi_- = \p_\zb \psib_- = 
\p_z \p_\zb \phi = \p_z \p_\zb \phib = 0
\eeq
imply 
\beq
&& \psi_+ = \psi_+(\zb), \quad \psi_- = \psi_-(z), \quad
\phi = \phi_+(\zb) + \phi_-(z) , \nnn
&& \psib_+ = \psib_+(\zb), \quad \psib_- = \psib_-(z), \quad
\phib = \phib_+(\zb) + \phib_-(z) .
\eeq
Consequently the supercurrents become
\beq
&& S_-(z) = \psi_-(z) \p_z \phib_-(z), 
\quad \bar S_-(z) = \psib_-(z) \p_z \phi_-(z),
\nnn && S_+(\zb) = \psi_+(\zb) \p_\zb \phib_+(\zb), 
\quad \bar S_+(\zb) = \psib_+(\zb) \p_\zb \phi_+(\zb).
\eeq
Not only is $S_-$ of left chirality, it is
left-moving, since it depends only
on $z$.  Similarly $S_+$ corresponds to
a right-moving superalgebra, depending only on $\zb$.  Thus, the
(2,2) notation corresponds equivalently to
counting the number of left- and right-moving 
supercurrents.  Similar remarks apply for
more general values (m,n).

\mys{Boundedness of the partition function}
\label{abre}
One might worry that due to the noncompact classical moduli
space of the undeformed theory, its partition function (for the
purely bosonic theory) is not well-defined.  Certainly this
is the case for case for the 0d reduction of $d \leq 4$ $SU(2)$ pure
Yang-Mills; however, this is not the case
for the 0d reduction of $d \geq 5$ $SU(2)$ pure
Yang-Mills; see Eq.~(24) of \cite{Krauth:1998xh}.  This is
because, properly speaking, the classical moduli space is a set
of measure zero in the field integration of the partition function.
For $d \geq 5$, ``entropic effects of the measure...overwhelm
the possible divergences'' \cite{Krauth:1998xh}.  

I am not aware of any exact result
showing that the partition function of the CKKU quiver system
obtained from the $6d \to 0d$ $SU(kN^2)$ pure Yang-Mills
is divergent.\footnote{The diagonal $U(1)$ from $U(kN^2) = U(1) 
\times SU(kN^2)$ decouples; its (divergent) contribution
to the partition function trivially factors out.}
In my view it is unlikely, just from the result
that $6d \to 0d$ $SU(2)$ pure Yang-Mills has a finite partition
function.  Furthermore, if there was a problem with the partition
function of the undeformed theory, I would have expected an
uncontrolled dispersion in the observable \myref{iwre} when I attempted
to measure it in the undeformed theory.  This is because
flat directions not suppressed by entropic effects would allow
for the Monte Carlo simulation to wander wildly 
throughout configuration space,
leading to results that would not converge to reliable average
values.  That this did not occur is further evidence that
the partition function is well-defined.  Finally, the deformation $S_{SB}$
can be regarded as a regulator of any possible divergence associated
with the noncompact flat directions.  I obtained stable, identical
results as the deformation was removed.  This provides further evidence that
the expectation values for the undeformed theory are reliable,
and that the quantum moduli space is compact.

\mys{Deformed fermion matrix}
\label{dfm}
Suppose we deform the fermion matrix $M$
according to $M_\e = M + \e \obf_{d_M}$,
where $d_M$ is the order (dimension) of
the fermion matrix $M$.  Then
if $M \psi = \lambda \psi$ determines
an eigenvalue $\lambda$ of $M$, it follows
that $M_\e \psi = \lambda_\e \psi$ with
$\lambda_\e = \lambda + \e$.  That is,
the set of eigenvectors of $M_\e$ is identical
to that of $M$; and, the set of eigenvalues of $M_\e$ is just
the set of eigenvalues of $M$ uniformly
shifted by $\e$.

I now consider $\e \to 0^+$.  Suppose that
$n_z=\dim \ker M$, the number of linearly independent
eigenvectors of $M$ with eigenvalues of zero.
Let $\det \hat M = \prod_{\lambda \not= 0} \lambda$
denote the product of nonzero eigenvalues
of $M$.  In the $\e \to 0^+$ limit, I note
\beq
\det M_\e = \prod_{\lambda_\e} \lambda_\e
= \e^{n_z} \prod_{\lambda \not= 0} \lambda
\( 1 + \frac{\e}{\lambda} \) = \e^{n_z}
\det \hat M \( 1 + \sum_{\lambda \not= 0} \frac{\e}{\lambda}
+ \ord{\e/\lambda}^2 \).
\eeq
Thus there exists a number $\alpha = 1 + \ord{\e/ |\lambda|_{{\rm min}}}$
such that
\beq
| \det M_\e - \e^{n_z} \det \hat M| \leq
\alpha (d_M - n_z) (\e/|\lambda|_{min}),
\eeq
where $|\lambda|_{min}$ is the minimum nonzero 
eigenvalue magnitude; i.e., $|\lambda|_{min} \equiv
\inf_{\lambda \not= 0} |\lambda|$.  It follows that
\beq
\lim_{\e \to 0^+} \frac{1}{\e^{n_z}} \det M_\e
= \det \hat M .
\label{kdm}
\eeq

In practice I take $n_z$ to be the number of
ever-present zeromodes.  Then through \myref{kdm}
I obtain the product of eigenvalues in the
subspace with these modes projected out.  This
is equivalent to having started with an $SU(kN^2)$
mother theory; for example, \myref{projl}
in the (2,2) 2d theory.

\mys{Conjugation of the action and properties of
the fermion determinant}
\label{scap}
\subsection{An example from kindergarten}
\label{aeki}
To begin, I discuss the simple example
of single-component Grassmann variables
$\psi, \psib$, with no site indices (i.e.,
a single-site lattice).  Let the
partition function have the simple form:
\beq
Z(M) = \int d\psib d\psi e^{-\psib M \psi},
\eeq
where $M$ is just a complex number.
Then according to the usual rules of Grassmann
integration, $Z(M) = M$.  Now define hermitian
conjugation to reverse the order of Grassmann variables
without the introduction of a sign, so that
\beq
(Z(M))^* = (Z(M))^\dagger =
\int d\psi^\dagger d\psib^\dagger e^{-\psi^\dagger M^* \psib^\dagger}.
\eeq
Note that the Grassmann variables are just dummy variables
of integration, so that we are free to replace
\beq
\psi^\dagger \to \psib, \quad \psib^\dagger \to \psi .
\label{trdf}
\eeq
Thus we find $(Z(M))^* = Z(M^*) = M^*$,
which is just what one should expect.  That is,
the rules of hermitian conjugation outlined
above are self-consistent.  
It may seem strange to be so very explicit
about this conjugation rule.  However, I
could take any other redefinition of the
integration variables and obtain the same result.
For example, instead of \myref{trdf} I could take
\beq
\psi^\dagger \to u \psib, \quad \psib^\dagger \to v \psi .
\label{trdg}
\eeq
Then using the Grassmann rules $d(u\psib) = (1/u) d\psib$, etc.,
\beq
(Z(M))^* = \frac{1}{uv} \int d\psib d\psi e^{-uv \psib M^* \psi}
= \frac{1}{uv} uv M^* = M^*.
\eeq
The rule \myref{trdf} is just a convenient
choice because it restores the measure and
the action to its original form, apart
from the factor $M$, which has been replaced
by $M^*$.  A more succinct description of
the generalized conjugation is as follows.
We define an operator $\Theta$ that is the composition
of hermitian conjugation and \myref{trdg}.
That is,
\beq
\Theta(\psi) = u \psib, \quad \Theta(\psib) = v \psi .
\eeq
$\Theta$ also has the property that it
conjugates complex numbers, transposes
matrix expressions and reverses the order
of fermions.  In more sophisticated examples
below, this freedom of choosing a $\Theta$ that defines
a generalized conjugation rule will become important.
In practice $\Theta$ that is unitary and satisfies
$\Theta^2=1$ is the most useful.

Everything that
follows will be a generalization of this
procedure, done in such a way as to remain
self-consistent in the above sense.
Note that if $M^* =M$, then $Z$ is also real.  Thus,
a hermitian action,
\beq
(\psib M \psi)^\dagger \equiv \psi^\dagger M^* \psib^\dagger
\to \psib M \psi,
\eeq
using \myref{trdf} and $M^*=M$, implies a real partition function $Z$.

\subsection{$n$-component complex Grassmann}
Here I will gather site indices and any
spin or internal indices into a collective
index $i=1,\ldots,n$.  The partition function is just
\beq
Z = \int d\psib_1 d\psi_1 \cdots d\psib_n d\psi_n
e^{-\psib M \psi},
\eeq
where $M$ is an $n \times n$ matrix.  As usual,
$Z(M) = \det M$.  Generalizing \myref{trdf} to
\beq
\psi_i^\dagger \to \psib_i, \quad \psib_i^\dagger \to \psi_i,
\label{nfhc}
\eeq
we find of course that
\beq
(Z(M))^* = (Z(M))^\dagger = Z(M^\dagger) = \det M^\dagger = (\det M)^*,
\eeq
which is right.  Thus if $M^\dagger = M$, one finds
that $Z(M)$ is real.  Using \myref{nfhc}, this
is equivalent to the statement that if the action
is self-conjugate, $Z(M)$ is real.

This self-conjugacy is sufficient but not necessary,
as I now show.  Suppose that instead of \myref{nfhc}
I take
\beq
\psi^\dagger \to \psib U, \quad \psib^\dagger \to V \psi,
\eeq
with $U,V$ $SU(n)$ matrices.  Then
\beq
Z^* &=& \int d\psi_n^\dagger d\psib_n^\dagger \cdots 
d\psi_1^\dagger d\psib_1^\dagger e^{-\psi^\dagger M^\dagger 
\psi^\dagger} \nnn
& \to & (\det U \det V)^{-1}
\int d\psib_1 d\psi_1 \cdots d\psib_n d\psi_n
e^{-\psib U M^\dagger V \psi}.
\eeq
Because $U,V$ are unimodular, the prefactor disappears.
Thus we find
\beq
\det M^* = Z(M)^* = \det (U M^\dagger V) .
\eeq
Suppose that we are able to find $U,V$ such that
\beq
U M^\dagger V = e^{i\varphi} M.
\label{umve}
\eeq
Then it follows that
\beq
\arg \det M = - \frac{n}{2} \varphi \mod \pi.
\label{argd}
\eeq

As in the 1-component case that was discussed in
\S\ref{aeki}, there exists a more concise formulation
of this result in terms of a generalized
conjugation operator $\Theta$.  That is, define
\beq
\Theta \psi = \psib U, \quad \Theta \psib = V \psi.
\eeq
Then taking into account \myref{umve},
\beq
\Theta(\psib M \psi) = \Theta \psi M^\dagger \Theta \psib
= \psib U M^\dagger V \psi = e^{i\varphi} \psib M \psi,
\eeq
and \myref{argd} follows.  In particular, if the
fermion action is self-conjugate in the generalized
sense that
\beq
\Theta (\psib M \psi) = \psib M \psi,
\eeq
it follows that $\det M$ is real.

\subsection{Imaginary-time continuation of hermiticity}
\label{itch}
The most common form of $\Theta$ is the
one that follows from hermiticity of the real-time
action.  To extract the imaginary-time continuation
of this conjugation operator, one just takes
$t_E = -it_M$, etc., in the real-time hermitian
conjugation rules.  For example, since hermitian
conjugation treats Minkowski time $t_M$ as a real variable,
it transforms Euclidean time as $t_E \to -t_E$.
An example is given in \S\ref{scpp} for the (2,2) model;
in particular, see \myref{econ}.  If the real-time
action is hermitian, then the imaginary-time action
is self-conjugate w.r.t.~this~$\Theta$.  The fermion
determinant is real in either description of
the Grassmann variables.

\mys{$\Ncal=4$ moduli space}
\label{n4ms}
Here I establish the well-known solution to \myref{qers}.
One way to see this is as follows \cite{Fayet:1978ig}.  First I note that
$S_0$ reduced to the zero modes, which I write as $S_z$,
takes the form
\beq
S_z &=& \frac{N^2}{g^2} \tr \( \half ([\xdag,x] + [\ydag,y] + [\zdag,z])^2 \right.
\nnn && \left. +2 [x,y][\ydag,\xdag] + 2[y,z][\zdag,\ydag] +2 [z,x][\xdag,\zdag] \) .
\label{jhas}
\eeq
Now note that the $U(1)$ parts of $x,y,z$ do not appear and can
take any value.  Thus I can restrict our attention the the $SU(k)$
parts, which I choose to express in terms of Hermitian matrices
$a_p,b_p, \; p=1,2,3$:
\beq
x^c T^c &=& (a_1^c + i b_1^c) T^c = a_1 + i b_1, \nnn
y^c T^c &=& (a_2^c + i b_2^c) T^c = a_2 + i b_2, \nnn
z^c T^c &=& (a_3^c + i b_3^c) T^c = a_3 + i b_3.
\label{ksfe}
\eeq
Substitution into \myref{jhas} and a bit of algebra yields
\beq
S_z &=& -\frac{N^2}{g^2} \tr \[ 2 \(\sum_p [a_p,b_p]\)^2
+ \sum_{p,q} \( [a_p, b_q] + [b_p, a_q] \)^2 \right. \nnn
&& \left. + \sum_{p,q} \( [a_p, a_q] - [b_p, b_q] \)^2 \] \nnn
&=& -\frac{N^2}{g^2} \sum_{p,q} \tr \( [a_p, a_q]^2
+ [b_p, b_q]^2 + 2 [a_p, b_q]^2 \).
\eeq
Using positivity arguments quite similar to those above,
one finds that $S_z \geq 0$  and that $S_z = 0$ iff
\beq
[a_p, a_q] = [b_p, b_q] = [a_p, b_q] = 0 , \quad \forall p,q,
\label{oiwe}
\eeq
which is nothing other than Eq.~(53) of \cite{Fayet:1978ig}.
Since the matrices are all hermitian and they all commute, 
it is obviously possible to choose a basis which simultaneously
diagonalizes them.  This basis will be related
to the one used in \myref{ksfe} according to 
$T^c \to T'^c = \alpha T^c \alpha^\dagger$,
which is nothing other than the global 
gauge transformations~\myref{oore}.

\end{document}